\pdfoutput=1
\documentclass[letterpaper, 10 pt, conference]{ieeeconf}  

\IEEEoverridecommandlockouts                              

\usepackage{graphics} 
\usepackage{epsfig} 
\usepackage{mathptmx} 
\usepackage{times} 
\usepackage{amsmath} 
\usepackage{amssymb}  
\usepackage{algorithm}
\usepackage{algorithmic}
\usepackage[utf8x]{inputenc} 
\usepackage{hyperref}
\usepackage{booktabs}
\usepackage{subcaption}
\usepackage{makecell}
\usepackage{multirow}
\newtheorem{remark}{\textbf{Remark}}
\newtheorem{definition}{\textbf{Definition}}
\newcommand{\RomanNumeralCaps}[1]
    {\MakeUppercase{\romannumeral #1}}

\title{\LARGE \bf
Driver Modeling through Deep Reinforcement Learning \\and Behavioral Game Theory
}

\author{Mert Albaba$^{1},\;Student\;Member$ and Yildiray Yildiz$^{2},\;Senior\;Member$
\thanks{$^{1}$Mert Albaba is with Faculty of Electrical and Electronics Engineering,
        Bilkent University, 06800 Bilkent, Ankara, Turkey
        {\tt\small mert.albaba@ug.bilkent.edu.tr}}%
\thanks{$^{2}$Yildiray Yildiz with the Department of Mechanical Engineering, 
		Bilkent University, 06800 Bilkent, Ankara, Turkey
        {\tt\small yyildiz@bilkent.edu.tr}}%
}

\begin{document}

\maketitle
\thispagestyle{empty}
\pagestyle{empty}

\begin{abstract}

In this paper, a synergistic combination of deep reinforcement learning and hierarchical game theory is proposed as a modeling framework for behavioral predictions of drivers in highway driving scenarios. The need for a modeling framework that can address multiple human-human and human-automation interactions, where all the agents can be modeled as decision makers simultaneously, is the main motivation behind this work. Such a modeling framework may be utilized for the validation and verification of autonomous vehicles: It is estimated that for an autonomous vehicle to reach the same safety level of cars with drivers, millions of miles of driving tests are required. The modeling framework presented in this paper may be used in a high-fidelity traffic simulator consisting of multiple human decision makers to reduce the time and effort spent for testing by allowing safe and quick assessment of self-driving algorithms. To demonstrate the fidelity of the proposed modeling framework, game theoretical driver models are compared with real human driver behavior patterns extracted from traffic data. 
\end{abstract}
\section{Introduction}

Autonomous vehicle (AV) technology is advancing rapidly along with the developments in decision-making algorithms. However, safety concerns about AV integration into daily traffic continue to exist, which needs to be addressed for a successful integration \cite{campbell2010autonomous}, \cite{anderson2014autonomous}. It is stated \cite{kalra2016driving} that millions of miles of on the road driving tests need to be conducted for a proper AV safety validation. In addition to real traffic tests, traffic environments simulated in computers may be used both to accelerate the validation phase and introduce a rich variety of traffic scenarios which may take several driver hours to encounter \cite{li2017game}, \cite{wongpiromsarn2009periodically}, \cite{lygeros1998verified}, \cite{wongpiromsarn2008formal}.  Although computer based traffic simulations offer easier and faster testing, to obtain reliable results, the agents (human driver models) in simulators should demonstrate human-like driving behavior at least with a reasonable accuracy.

Several approaches are proposed in the open literature to obtain high fidelity human driver models. Markov Dynamic Models (MDMs) are utilized for the prediction and recognition of the driver maneuvers, such as steering or braking, in \cite{liu2001modeling}. An approach named SITRAS (Simulation of Intelligent Transport Systems) is proposed in \cite{hidas2002modelling}, for modeling lane changing and merging behaviors. Dynamic Bayesian Networks (DBNs) are presented in \cite{dagli2002action}, for recognition of acceleration or lane changing actions. In \cite{ungoren2005adaptive} adaptive predictive control is proposed for modeling drivers' lateral actions. Cognitive architecture method, which is a framework that specifies computational behavioral models of human cognition is presented in \cite{salvucci2006modeling} for steering and lateral position modeling. A combination of cognitive psychology and artificial intelligence methods, is utilized in \cite{liu2007human} for predicting behaviors of human drivers at signalized and unsignalized intersections. In \cite{kumar2013learning}, support vector machine (SVM) with Bayesian filter is utilized for predicting lane change intentions of drivers. In \cite{aoude2013probabilistically} and \cite{tran2013modelling}, Gaussian Process Regression is employed to predict driver behaviors and detect behavioral patterns. For the estimation of driver decisions at intersections, a framework which models vehicle dynamics and driver behaviors as a hybrid-state system is proposed in \cite{gadepally2014framework}. In order to obtain driving styles from vehicle trajectories, inverse reinforcement learning is used in \cite{kuderer2015learning}. Hidden Markov Model based driver models are derived in \cite{lefevre2015autonomous}. In \cite{burton2016driver}, SVMs are used to model driving patterns. A neural network based method, which utilizes recurrent neural networks, is presented in \cite{morton2016analysis} for probabilistic driving modeling. An SVM based model is given in \cite{zhao2017modeling} for predicting driving behavior at roundabouts. For modeling human highway driving behavior, Generative Adversarial Imitation Learning is proposed in \cite{kuefler2017imitating}. Human drivers' stopping behaviors are modeled in \cite{da2018biologically}. Also, in \cite{han2019human}, an optimal control theory based approach is presented to model stopping behavior. Gaussian mixture model and hidden Markov model are combined in \cite{wang2018learning} for the prediction of lateral trajectories. A car-following driver model imitating human driving is proposed in \cite{hu2020car}, using a neural network based modeling approach. In \cite{sadigh2016planning}, interactions between an autonomous vehicle and a human driver is modeled. Apart from individual driver models, traffic flow models are also studied. Some examples can be found in \cite{li2004modeling}, \cite{wang2011new} and \cite{abadi2015traffic}.

In this paper, we propose a deep reinforcement learning and game theory based stochastic modeling method which allows simultaneous decision making for multi agent traffic scenarios. What distinguishes our method from existing studies is that all the drivers in a multi-move scenario make strategic decisions simultaneously, instead of modeling the ego driver as a decision maker and assuming predetermined actions for the rest of the drivers. This is achieved by combining a hierarchical game theoretical concept named \textit{level-k reasoning} with a reinforcement learning method called \textit{Deep Q-learning} (DQN) \cite{mnih2013playing}. There exist earlier studies that also use reinforcement learning and game theory in modeling driver behavior, such as \cite{albaba2019modeling}, \cite{albaba2019stochastic} \cite{li2017game}, \cite{li2016hierarchical} and \cite{oyler2016game}. The contribution of this work over these earlier results can be listed below: 
\begin{enumerate}
\item In this study, the proposed method can model a dramatically larger class of scenarios which was not possible earlier. This is achieved thanks to the exploitation of DQN, where the enlarged state space is handled by a deep neural network;
\item Unlike earlier results, crash rates in the simulations are reduced to their real life realistic levels, by eliminating the driver blind spots, with an enlarged observation space;
\item Developed driver models are compared with two different sets of traffic data to provide a statistical analysis of model prediction capability.
\end{enumerate}

There are other game theoretical driver models proposed in the literature. For example, in \cite{yoo2012stackelberg}, a Stackelberg game structure is used to model highway driving. However, dynamic scenarios consisting of several moves are not considered. Another work, presented in \cite{dextreit2013game}, which also uses Stackelberg games, considers multi-move scenarios. However, computations become quite complex once the number of players (drivers) increases to more than 2, especially for dynamic scenarios.

This paper is organized as follows. In Section \RomanNumeralCaps{2}, deep reinforcement learning methods, level-k reasoning and the algorithm combining these methods are described. In Section \RomanNumeralCaps{3}, vehicle physical models, raw traffic data processing, and driver observation and action spaces are explained along with the reward function. In Section \RomanNumeralCaps{4}, creation of strategic agents is explained. In Section \RomanNumeralCaps{5}, the statistical comparison method employed for comparing model behavior with data, the comparison procedure, crash rate analysis and the statistical model-data comparison results are given. Finally, a summary is provided in Section \RomanNumeralCaps{6}. 

\section{Method}
\subsection{Level-k Reasoning}

\begin{figure}[h!]
	\centering
	\includegraphics[width=\linewidth]{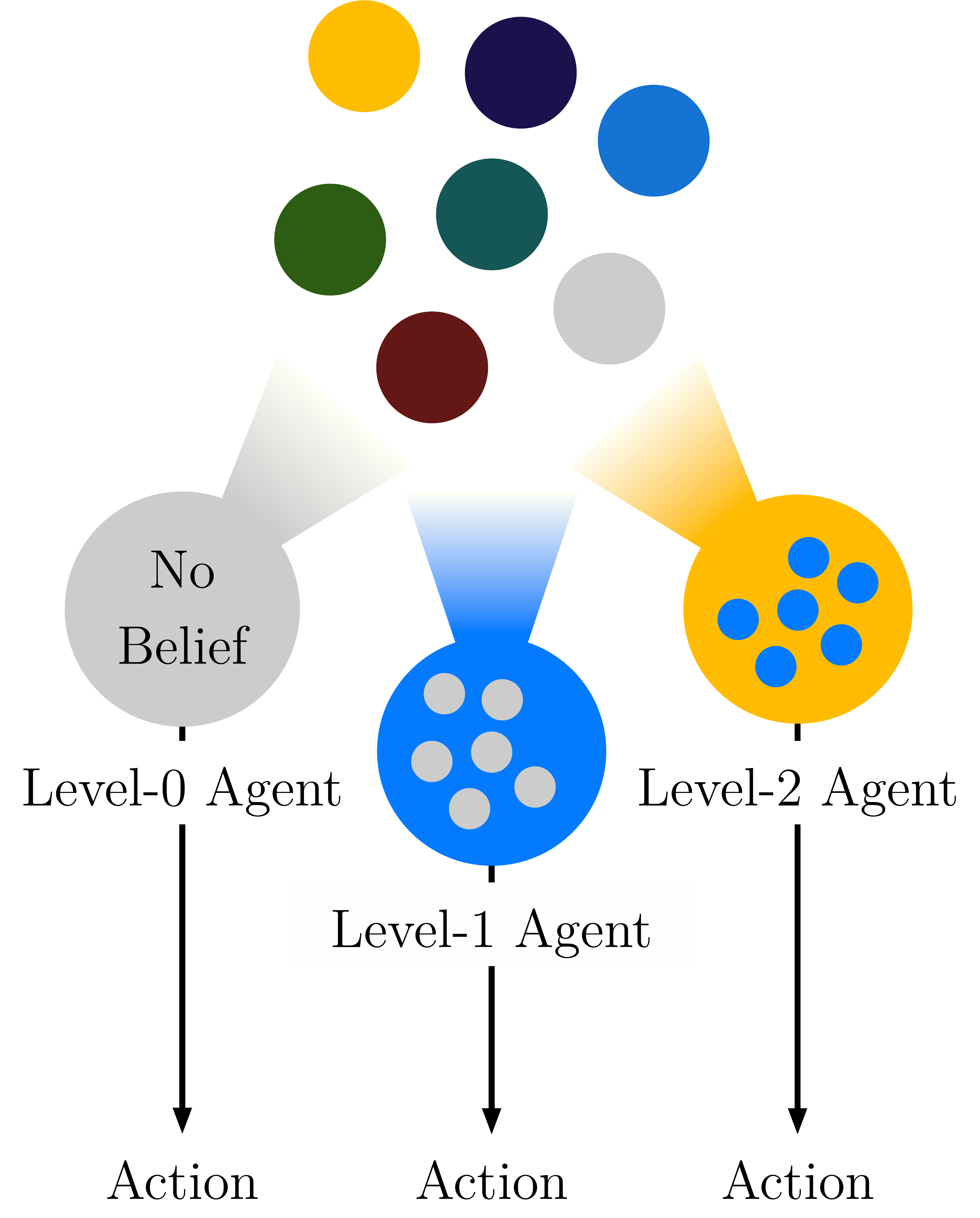}
	\caption{A level-0 agent (grey) does not consider other agents' possible moves before making a decision; a level-1 agent (blue) assumes that all other agents are level-0 and takes action based on this assumption; a level-2 agent (yellow) believes that other agents are level-1 and acts accordingly.}
	\label{levelk}
\end{figure}

In order to model strategic decision making process of human drivers, a game theoretical concept named  \textit{level-k reasoning} is used \cite{camerer2004cognitive}, \cite{nagel1995unraveling}, \cite{stahl1995players}. Level-k approach is a hierarchical decision making concept and presumes that different levels of reasoning exist for different humans. The lowest level of reasoning in this concept is called \textit{level-0 reasoning}. A level-0 agent is a non strategic/naive agent since his/her decisions are not based on other agents' possible actions, but consists of predetermined moves, which may or may not be stochastic. In one level higher, a strategic level-1 agent exists, who determines his/her actions by assuming that the other agents' reasoning levels are level-0. Hence, actions of a level-1 agent are the best responses to level-0 actions. Similarly, a level-2 agent considers other agents as level-1 and makes his/her decisions according to this prediction. This process continues following the same logic for higher levels. In Fig. \ref{levelk} the general architecture of level-k thinking is presented. The figure shows a population consisting of level-0 (gray), level-1(blue) and level-2(yellow) agents. In this population, level-0's are non-strategic, meaning that they do not consider other agents' possible actions before they make their own move. Strategic level-1 and level-2 agents, on the other hand, assumes that everyone else is level-0 and level-1, respectively, and act accordingly. In some experiments, humans are observed to have at most level-3 reasoning \cite{costa2009comparing}, which may, of course, depend on the type of the game being played. To generalize, all level-k agents, except level-0, presume that the rest of the agents are level-(k-1) and makes their decisions based on this belief. Since this belief may not always hold true, the agents have bounded rationality.

\begin{remark}
A level-k agent performs optimally against a level-(k-1) agent, however, may not perform optimally against other levels, which introduces bounded rationality to agent actions. 
\end{remark}

\subsection{Deep Q-Learning}
In the previous section, the exploitation of level-k reasoning to model strategic decision making is explained. In time-extended scenarios, where the agents make a series of decisions before an episode is completed, such as the traffic scenarios focused on in this paper, level-k reasoning can not be used alone. To obtain driver models that provide best responses to the other agents' likely actions in a multi-move setting, we utilize Deep Q-Learning (DQN) together with level-k reasoning. The main reason for the employment of DQN is the large state space that becomes infeasible to handle with other reinforcement learning (RL) methods used in earlier studies \cite{albaba2019modeling}, \cite{li2017game}, \cite{li2016hierarchical}, \cite{oyler2016game}. In this section, a brief description of DQN tailored for the task at hand is given. More detailed expositions of DQN can be found at \cite{mnih2013playing}, \cite{lample2017playing} and \cite{roderick2017implementing}. 

\begin{figure}[h!]
	\centering
	\includegraphics[width=\linewidth]{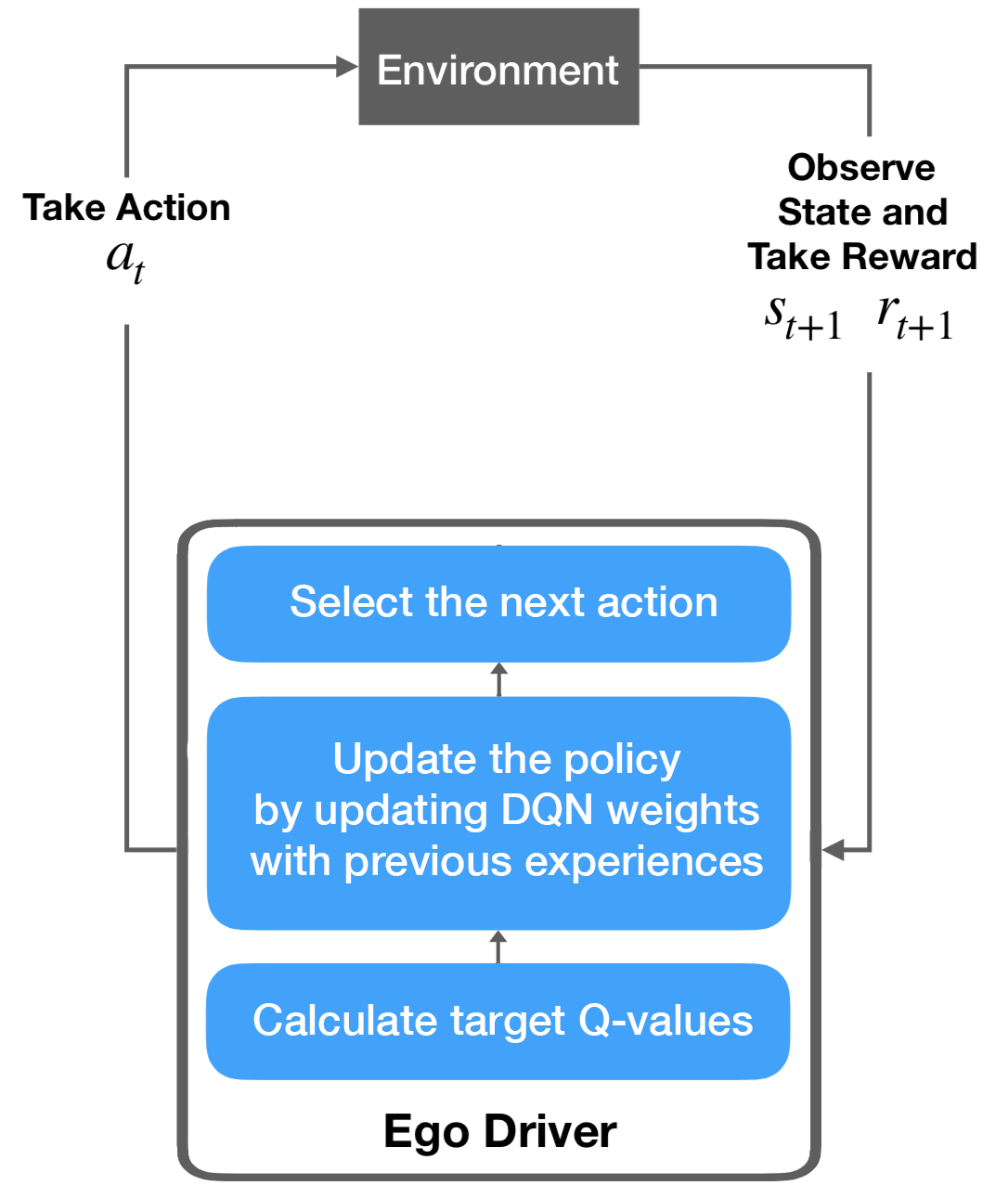}
	\caption{Reinforcement Learning Process Diagram}
	\label{diagram}
\end{figure}

RL is a learning process through reward and punishment \cite{sutton1998reinforcement}. At each step of learning, or training, the trained agent observes a state $s_t \in S$ (state $s$ at time $t$), selects an action $a_t \in A$ (action $a$ at time $t$), and as a result, transitions into a new state $s_{t+1} \in S$ with a certain probability. The sets $S$ and $A$ represent observation and action spaces, respectively. The agent receives a reward after the transition based on a reward function, which is a mathematical expression of the preferences of the agent. This process, i.e. observation, action, transition and reward accumulation, continues until the average reward converges to a certain value. The process is depicted in Fig. \ref{diagram}. The main goal of the agent is selecting actions that maximizes the total obtained rewards, where future rewards may have decreasing levels of importance in the agent's decision making and thus, can be ``discounted". In other words, the agent tries to find a policy  $ \pi: \textit{S} \longrightarrow \textit{A} $, a mapping from states to actions, which maximizes the expected discounted cumulative reward. This policy is called the optimal policy: $ \pi^{*} $. In RL, value function $V(s)$ estimates the cumulative reward starting from state $s$. It is a measure of the value of being in state $s$. Action value function, $Q(s,a)$, estimates the cumulative reward obtained by the agent by starting with action $a$ in state $s$. It represents how valuable selecting action $a$ in state $s$ is.

An agent's value function, or expected cumulative discounted reward when starting from state $s$, and following a policy $\pi$ is defined as

\begin{equation}
V^{\pi}(s) = E(\sum_{t\geq 0}\gamma ^tr_t),
\end{equation}

\noindent for all $s \in S$, where $\gamma$ represents a discount factor, $r_t$ represents the reward at time-step t and $s$ represents the observed state. The corresponding optimal value function is given as

\begin{equation}
V^{*}(s) = \max_{\pi }V^{\pi}(s).
\end{equation}

The optimal value function is related to the optimal Q-function, $Q^{*}$, as
\begin{equation}
V^{*}(s) = \max_{a}Q^{*}(s,a).
\end{equation}
Moreover, the optimal policy can be defined in terms of the optimal Q-function as

\begin{equation}
\pi^{*}(s) = arg\max_{a} Q^{*}(s,a).
\end{equation}

\noindent Bellman equation provides a relationship between the value of a state-action pair \textit{(s,a)} and its successor pairs through

\begin{equation}
Q(s,a) = E[r_{t+1} + \gamma\max_{a^{'}}Q(s_{t+1}, a^{'} )\; \vert \; s_{t} = s, a_{t} = a],
\label{bellmaneqn}
\end{equation}

\noindent where $E$ represents the expected value and $(s_{t+1}, a^{'})$ is the next state-action pair. \textit{Q-learning} algorithm provides a method, based on the Bellman Equation (\ref{bellmaneqn}), to update action values iteratively, which is given by

\begin{equation}
\begin{aligned}
Q_{t+1}(s_t,a_t) = Q_t(s_t,a_t)+\alpha (r_{t+1}+\\ \gamma \max_{a}Q_t(s_{t+1},a)-Q_t(s_t,a_t)),
\end{aligned}
\end{equation}

\begin{figure}[!t]
\centering
  \includegraphics[width=1.75in]{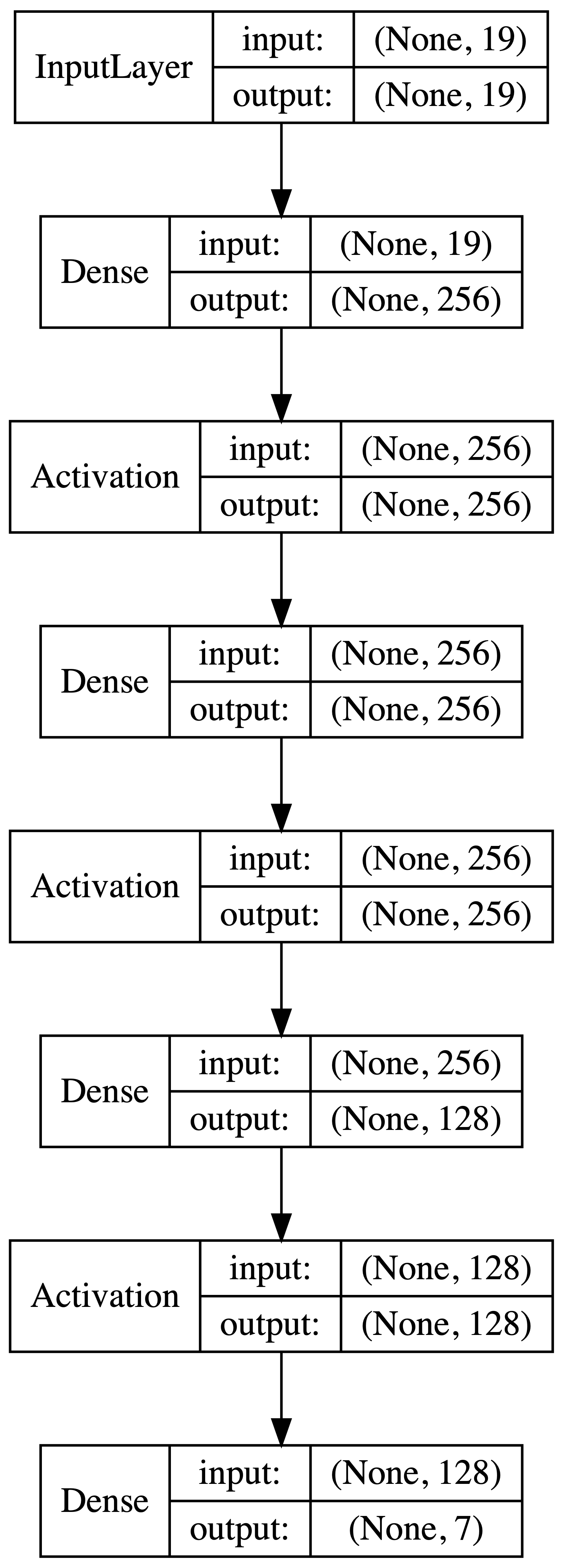}
  \caption{A deep neural network consisting 4 layers is utilized to approximate the Q-function. The first layer, \textit{InputLayer} takes state representations as inputs. \textit{Dense} layers refers to a fully connected neural network layer, i.e. there is a connection between each input feature and each node of the layer. Activation layers implements activation functions, which introduces non-linearity. There are none values in each shape since different batch sizes are accepted by the network, i.e. none values are automatically replaced by given batch size. }
  \label{neural}
\end{figure}

\noindent where $\alpha$ is a learning rate and $\gamma$ is the discount factor. Q-learning is guaranteed to converge if each state-action pairs are visited infinitely many times during training \cite{sutton1998reinforcement}.

Keeping a table containing action values for all state-action pairs may be infeasible for large state spaces. In these cases, a neural network function approximator with a weight vector $\theta$ can be used to approximate the action-value function $Q(s,a) \approx Q(s,a; \theta )$ \cite{lample2017playing}. In Deep Q-Learning (DQN), values of Q-function is approximated with a deep neural network, which proves useful especially for large state spaces. Neural network architecture utilized in this work is given in Fig. \ref{neural}. 

DQN stores last $N$ experiences in memory $D$. Each experience is a four-tuple: $(s, a, s^{'} , r)$, where $a$ is the action made in state $s$, $s^{'}$ is the transitioned state after taking the action $a$, and $r$ is the obtained reward. In the first $n_s$ steps of training, network weights are not updated. Starting from the $n_{s+1}^{st}$ step, at each step, a  mini-batch of the stored four-tuples  is randomly sampled from memory $D$ and the Q-function is updated, through network weight updates.

\begin{algorithm}[!t]
	\caption{Deep Q-Learning}
	\begin{algorithmic}[1]
		\STATE Initialize the memory \textit{D} to capacity \textit{N}
		\STATE Initialize the main network and the target network, $QN$ and $QN^{T}$, with weights sampled from a uniform distribution of range $[-\sqrt{\dfrac{6}{n_{input}+n_{output}}},\sqrt{\dfrac{6}{n_{input}+n_{output}}}]$, where $n_{input}$ and $n_{output}$ are number of input and output neurons, respectively \cite{glorot2010understanding}. 
		\STATE Set $T=50$
		\FOR {episode = 1 to $M$} 
			\FOR {t = 1 to $K$}
				\STATE Sample action $a_t$ using the probability values $P_t(a_i) = \dfrac{e^{QN_t(a_i)/T}}{ \sum_{j=0}^{n-1} e^{QN_t(a_j)/T}}$, $i=1,2, ... , size(Action Space)$
				\STATE Execute action $ a_t $ and observe the reward $ r_t $ and the transitioned state $s_{t+1}$
				\STATE Store the experience $(s_t,a_t,r_t,s_{t+1})$ in $D$
				\IF {size($D$) $\geq$ $n_s$}
					\STATE Sample a random batch of experiences, consisting of P four-tuples $(s_j,a_j,r_j,s_{j+1})$
					\FOR {j = 1 to $P$}
						\STATE Set $y_j = r_j + \gamma \max_{a^{'}}QN^{T}(s_{j+1},a^{'}; W)$
						\IF {$s_{j+1}$ is terminal, i.e. ego driver crashes,}
							\STATE Set $y_j = r_j$
						\ENDIF
						\STATE Perform a gradient descent step using the cost function$(y_j-NN(s_{j+1},a_j;W))^2$ with respect to weight matrix $W$
					\ENDFOR
				\ENDIF	
				\IF  {$s_{t+1}$ is terminal, i.e. ego driver crashes,}
					\STATE break
				\ENDIF
			\ENDFOR 
			\IF  {$T > 1$}
					\STATE Update Boltzmann Temperature $T = T*c$, $c<1$
			\ENDIF	
		\ENDFOR
	\end{algorithmic}
\end{algorithm}

Boltzmann exploration, i.e. softmax, with initial temperature of $T=50$, which decreases exponentially to 1 during the training process, is used to obtain a stochastic neural network instead of a deterministic one. In Boltzmann exploration, during training, probabilities of actions are calculated with the Boltzmann distribution and actions are taken probabilistically. At time-step $t$, probability of taking action $a$ is
\begin{equation}
P_t(a) = \dfrac{e^{Q_t(a)/T}}{\sum_{i=0}^{n-1} e^{Q_t(a_i)/T}},
\end{equation}
where $T$ is the temperature and $n$ represents the number of actions \cite{ravichandiran2018hands}. When the temperature is high, the probability of the actions are closer to each other. When the temperature is low, the probability of the action with highest Q-value is higher than the rest of the actions. 

Algorithm pseudo-code for DQN is given in Algorithm 1, where $M$ is the total number of training episodes, $K$ is the number of steps within an episode, $P$ is the number of four-tuples in a mini-batch, and $T$ is the temperature in Boltzmann distribution.

\subsection{Combining Level-k Reasoning with Deep Q Learning}

To generate agents with different levels of reasoning for modeling multi-move strategic decision making in traffic scenarios, learning capability offered by DQN is combined with the level-k reasoning approach.

In the proposed combined approach, the predetermined, non-strategic level-0 policy is the anchoring policy from which all the higher levels are derived using Deep Q-learning (DQN). For example, in order to obtain the level-1 policy, a traffic scenario is created where all drivers are level-0 agents except the \textit{ego driver},  who is to be trained via DQN to learn how to best respond to the level-0 policy. The details of this training is given in Section II-B. Once the training is over, the ego driver becomes a level-1 agent. The procedure for obtaining the level-1 policy through proposed combination of level-k reasoning and DQN is explained in Algorithm \ref{alg:L1_TR}, where $n_d$ is number of drivers.

\begin{algorithm}
	\caption{Obtaining the level-1 policy by combining DQN and Level-k Reasoning}
	\label{alg:L1_TR}
	\begin{algorithmic}[1]
		\STATE Load the predetermined level-0 policy, $\pi^0$
		\STATE Set the reasoning levels of all agents in the environment, $p_i$, to level-0: $\pi_{p_i} = \pi^0,$ $i = 1,2,3,...,n_d$ 
		\STATE Initialize the ego driver's policy to a uniform action probability distribution over all states: $\pi_{ego} = \pi^{uniform}$
		\STATE Train the ego driver using DQN
		\STATE At the end of the training, ego driver learns to best respond to $\pi^0$, and therefore the resulting policy is the level-1 policy, $\pi^1$.
	\end{algorithmic}
\end{algorithm}

The level-2 policy can be obtained similarly: A level-2 player assumes that all other players in the environment are level-1, and takes actions based on this assumption, therefore his/her actions are best responses to level-1 players. For the training of the level-2 agent, a traffic scenario where all the agents, except the ego driver, are assigned the previously obtained level-1 policy. Using DQN, the ego driver is trained to give the best responses to level-1 drivers, hence the resulting policy becomes the level-2 policy. This process is given in Algorithm \ref{alg:L2_TR}

\begin{algorithm}
	\caption{Training of the level-2 agent by combining DQN and Level-k Reasoning}
	\label{alg:L2_TR}
	\begin{algorithmic}[1]
		\STATE Load the previously obtained level-1 policy, $\pi^1$ (see Algorithm \ref{alg:L1_TR})
		\STATE Set the agents in the environment, $p_i$, as level-1 agents: $\pi_{p_i} = \pi^1,$ $i = 1,2,3,...,n_d$ 
		\STATE Initialize the ego driver's policy to a uniform action probability distribution over all states: $\pi_{ego} = \pi^{uniform}$
		\STATE Train the ego driver using DQN
		\STATE At the end of the training, ego driver learns to best respond to level-1 policy, $\pi^1$. Thus, the resulting policy is the level-2 policy, $\pi^2$.
	\end{algorithmic}
\end{algorithm}

\begin{algorithm}
	\caption{Training of the level-k agent by combining RL and Level-k Reasoning}
	\label{alg:LK_TR}
	\begin{algorithmic}[1]
		\STATE Load the level-(k-1) policy, $\pi_{k-1}$
		\STATE Assign level-(k-1) policy to the agents in the environment, $p_i$, as:  $\pi_{p_i} = \pi^{k-1}$ 
		\STATE Place level-(k-1) agents in the environment
		\STATE Set empty policy to the learning agent: $\pi_{ego} = \pi^{empty}$
		\STATE Train the learning agent with DQN
		\STATE At the end trained agent learns best responses to the level-(k-1), so save the resulting policy as the level-k policy, $\pi_{k}$
	\end{algorithmic}
\end{algorithm}

\begin{figure*}[h!]
	\centering
	\includegraphics[width=\linewidth]{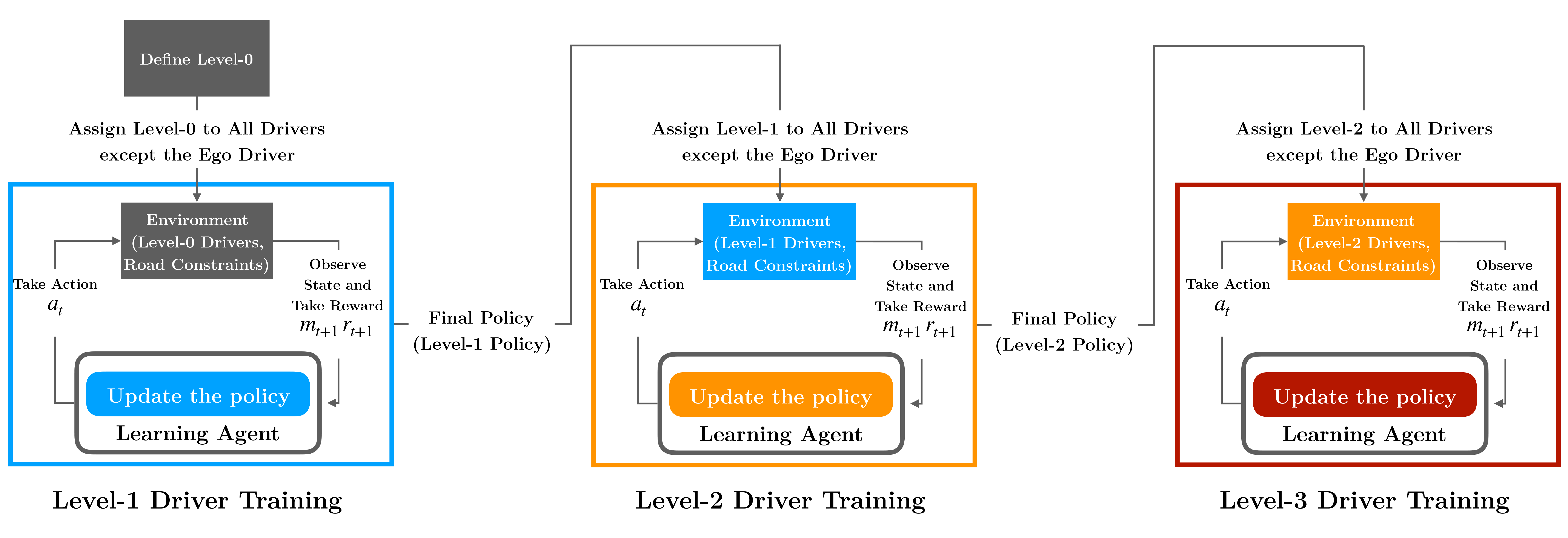}
	\caption{Combination of Level-k reasoning and reinforcement learning.}
	\label{trainprocess}
\end{figure*}

Higher level policies can also be obtained in a similar way, if desired. Training process for obtaining a level-k policy is described in Algorithm \ref{alg:LK_TR}. It is noted that the hierarchical learning process explained above decreases the computational cost since at each stage of learning, the agents other than the ego agent use previously trained policies and hence become parts of the environment. This helps to obtain traffic scenarios, containing a mixture of different levels, where all the agents are simultaneously making strategic decisions. This sharply contrasts conventional driver decision making approaches, in a crowded traffic, where one or two drivers are strategic decision makers and the rest are assigned predefined policies that satisfy certain kinematic constraints. In this work, highest level is set to level-3, in accordance with \cite{costa2009comparing}.  A visual representation of the process of combining level-k reasoning and DQN is given in Fig. \ref{trainprocess} 

\section{Traffic Scenario}
The traffic scenario that is used to train and test driving policies comprise of a five lane highway and multiple vehicles. The lane width of the highway is 3.7m and the size of each vehicle is 5m x 2m. The vehicles have continuous dynamics. In the following subsections, the elements of the traffic scenario are explained. Certain numerical values that are needed to create the traffic scenario, such as observation and action space parameters, are determined based on one of the two sets of traffic data. Therefore, we first explain how data is processed before providing the scenario details. It is noted that in this section we only describe data processing for use in determining specific traffic scenario parameters. Processing of the data to obtain real human driver policies is explained later in Section IV.

\subsection{Traffic Data Processing}
In this work, two sets of traffic data, collected on US101 and I80 highways \cite{US101}, \cite{I80}, are used for model validation. Among these two, the US101 set is employed to determine the observation and action space parameter values. These data sets consist of time and vehicle ID stamped position, velocity and acceleration values. Before employing the data, firstly, inconsistencies in the acceleration and velocity values are addressed: A careful analysis demonstrates some contradictions in acceleration values and large velocity jumps over consecutive time steps in the traffic data. The problem of large velocity jumps is solved by applying a linear curve fitting. To exemplify, if among the velocity values $v_{i-5}$, $v_{i-4}$, ..., $v_{i}$, $v_{i+1}$, $v_{i+2}$, ..., $v_{i+5}$, where the subscripts denote the time steps, the values $v_{i+1}$ and $v_{i+2}$ showed impossible jumps, these values are replaced with appropriate values $v_{i+1}=v_{i}+\dfrac{v_{i+3}-v_{i}}{3}$ and $v_{i+2}=v_{i}+\dfrac{2(v_{i+3}-v_{i})}{3}$. Once unreasonable velocity jumps were eliminated, acceleration values are obtained by using the five-point stencil method \cite{sauer2017numerical}, \cite{abramowitz1948handbook} given as

\begin{equation}
\label{numdif}
\begin{split}
a_i = \dfrac{-v_{i+2\delta i} + 8v_{i+\delta i}}{12\delta i}\\ 
+  \dfrac{-8v_{i-\delta i} + v_{i-2\delta i}}{12\delta i}.
\end{split}
\end{equation}

\noindent Since $v_{i+2\delta i} $, $v_{i+\delta i}$, $v_{i-\delta i}$ and $v_{i-2\delta i}$ do not exist for the first and the last two time-steps for each car, 5-point finite difference method is used to calculate the acceleration values for these points. Using this method, the first and last time step accelerations are calculated as

\begin{equation}
\begin{split}
a_i = \dfrac{-25v_i + 48v_{i+\delta i}-36v_{i+2\delta i}}{12\delta i} \\
+ \dfrac{16v_{i+3\delta i}-3v_{i+4\delta i}}{12\delta i}
\end{split}
\end{equation}

\begin{equation}
\begin{split}
a_i = \dfrac{3v_{i-4\delta i} - 16v_{i-3\delta i}+36v_{i-2\delta i}}{12\delta i} \\
+ \dfrac{- 48v_{i-\delta i}+25v_i}{12\delta i},
\end{split}
\end{equation}

\noindent respectively. Similarly, for the second and second-to-last time steps, the accelerations are calculated as

\begin{equation}
\begin{split}
a_i = \dfrac{-3v_{i-\delta i} - 10v_i+18v_{i+\delta i}}{12\delta i} \\
+ \dfrac{ - 6v_{i+2\delta i}+v_{i+3\delta i}}{12\delta i}
\end{split}
\end{equation}

\begin{equation}
\begin{split}
a_i = \dfrac{-v_{i-3\delta i} + 6v_{i-2\delta i}-18v_{i-\delta i}}{12\delta i} \\
 + \dfrac{10v_i+3v_{i+\delta i}}{12\delta i},
\end{split}
\end{equation}

\noindent respectively.

Both US101 Data and I80 Data consists more than 5 lanes. Because of that, in order to make them comparable to our model, which is for 5-lane road, we have decreased the number of lanes by considering cars on lanes, which are right to the $5^{th}$, as on $5^{th}$ lane. 

\subsection{Driver Observation Space}
In traffic, drivers cannot observe all the cars on the road, but observe the cars around their vicinity. In this work, it is assumed that a driver on lane $l$ observes the closest front and rear cars on lanes $l-2, l-1, l+1$ and $l+2$ along with the front car on lane $l$. Therefore, up to 9 surrounding cars are observable by the driver. Observations are coded as relative positions and velocities. Specifically, a driver on lane $l$ can detect

\begin{itemize}
\item Relative position and velocity of the car in front on the same lane (lane $l$).
\item Relative position and velocity of the car in front on the left lane (lane $l+1$).
\item Relative position and velocity of the car in rear on the left lane (lane $l+1$).
\item Relative position and velocity of the car in front on the right lane (lane $l-1$).
\item Relative position and velocity of the car in rear on the right lane (lane $l-1$).
\item Relative position and velocity of the car in front on two left lane (lane $l+2$).
\item Relative position and velocity of the car in rear on two left lane of (lane $l+2$).
\item Relative position and velocity of the car in front on two right lane (lane $l-2$).
\item Relative position and velocity of the car in rear on two right lane (lane $l-2$).
\item Own lane number ($l$).
\end{itemize}

In this work, both continuous and discrete observation spaces are used for modeling, separately. When we use the discrete observation space, we name the method as regular DQN. When we use continuous observation space we name the method as c-DQN. We provide data validation results for both of the cases. It is noted that regular DQN is computationally less expensive but c-DQN is more accurate. Below we provide the details of both DQN and c-DQN.

\subsubsection{DQN}
Human driver observations, consisting of relative positions, $\delta x$, and velocities, $\delta v$, of the surrounding vehicles, are quantized into different sets: Relative positions are binned as \textit{close}, \textit{nominal} and \textit{far}, while the bins used for relative velocities are \textit{stable}, \textit{approaching} and \textit{moving away}. In order to determine the contents of these bins, the raw US101 traffic data \cite{US101} is processed. Below, we explain how the relative position and relative velocity bins are determined. 

\begin{figure}[h!]
	\centering
	\includegraphics[width=\linewidth]{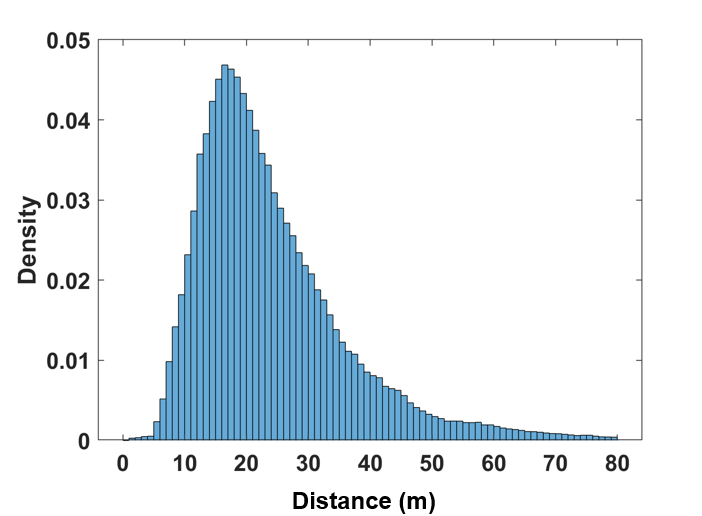}
	  \caption{Distribution of distances between consecutive vehicles presented as a probability density function.}
	\label{distances}
\end{figure}

The distribution of distances between vehicles, obtained by processing US101 traffic data, is shown in Fig. \ref{distances}. It is seen that around 50\% of the time, the distance between vehicles stays between 11m and 27m. Based on this observation, the driver relative position observations are binned as \textit{close}, if $\delta x < 11\text{m}$, \textit{nominal} if $11\text{m}< \delta x <27\text{m}$, and \textit{far} if $\delta x > 27\text{m}$.   

Drivers also observe the changes in distances to cars around them, which can be thought in terms of relative velocities, $\delta v$. In DQN, we bin the relative velocities as approaching if $\delta v<-0.1\text{m/s}$, stable if $-0.1\text{m/s} <\delta v < 0.1\text{m/s}$ and moving away if $\delta v >0.1\text{m/s}$.

As a result, the ego driver observation bins, for a single surrounding vehicle,  consist of 3 relative position bins and 3 relative velocity bins. Furthermore, he/she can observe his/her own lane number (1-5). Considering 9 surrounding vehicles, the size of the observation space equals to $3^9*3^9*5 = 3^{18}*5 = 1937102445$, which approximately equals to 2 billion. Hence, even after binning the observations, the resulting observation space size is quite large.

\begin{remark}
In earlier studies \cite{albaba2019modeling}, \cite{li2017game}, \cite{li2016hierarchical}, \cite{oyler2016game}, at most 5 surrounding vehicles were included in the observation space, which made its size at most $295245$. In this work, thanks to the DQN method, which was not employed earlier, a dramatically larger observation space can be handled. 
\end{remark}

\subsubsection{c-DQN} 
The main motivation behind using binned observations, which is the case for regular DQN explained above, is simplifying the learning process without introducing unreasonable assumptions. However, reduced observation resolution may result in inaccurate decisions in critical conditions. To address this problem, we tested the c-DQN approach, where continuous observations are used without any binning. This approach provides an accurate view of the surrounding vehicles to the learning process but makes the observation space infinitely large.

\begin{figure}[h!]
  \includegraphics[width=\linewidth]{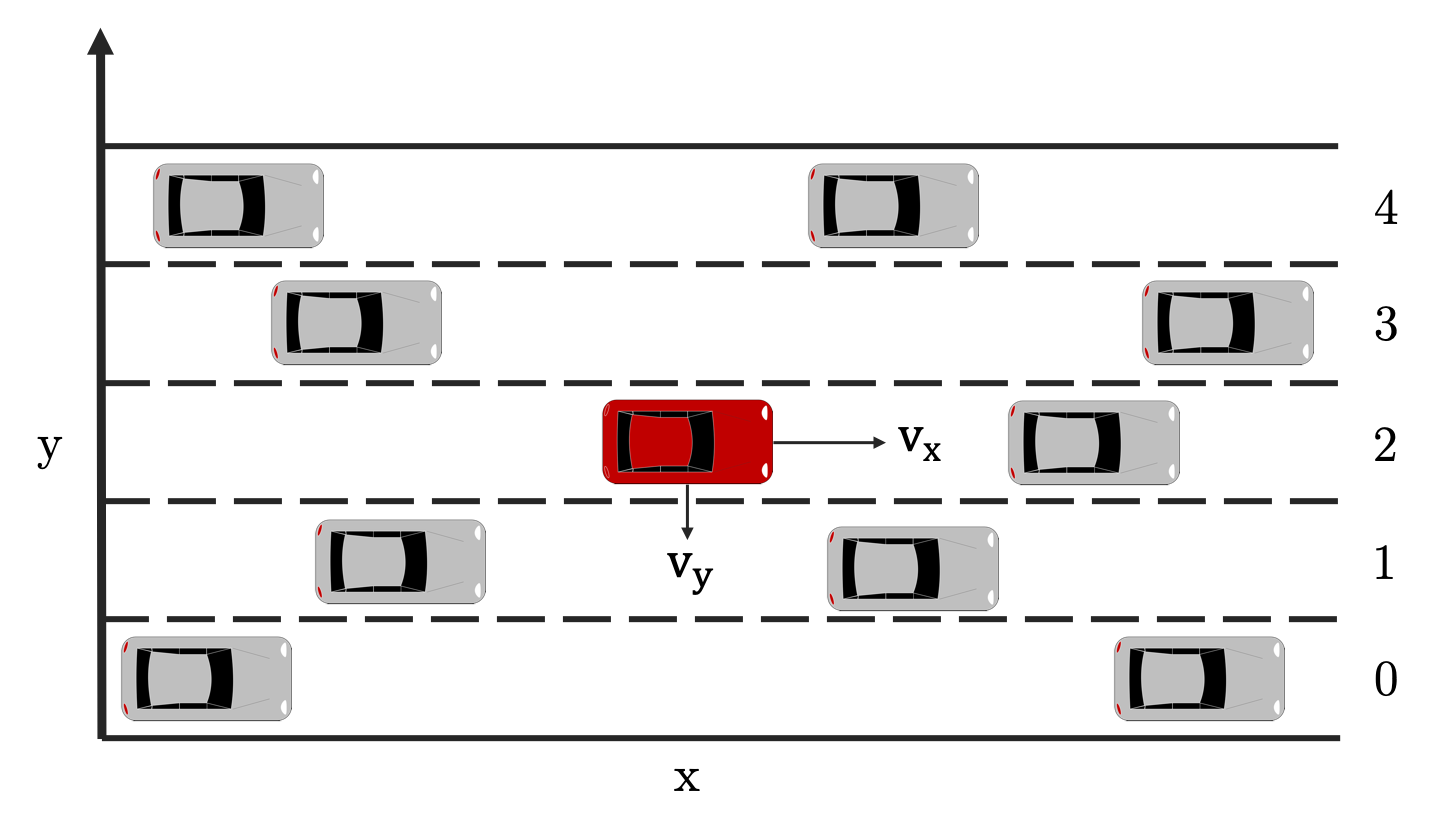}
  \caption{The ego vehicle (red, center) and the vehicles the ego driver can observe. Lane numbers are shown on the right. }
  \label{fig:model1}
\end{figure}

\subsection{Driver Action Space}
\begin{figure}[h!]
	\centering
	\includegraphics[width=\linewidth]{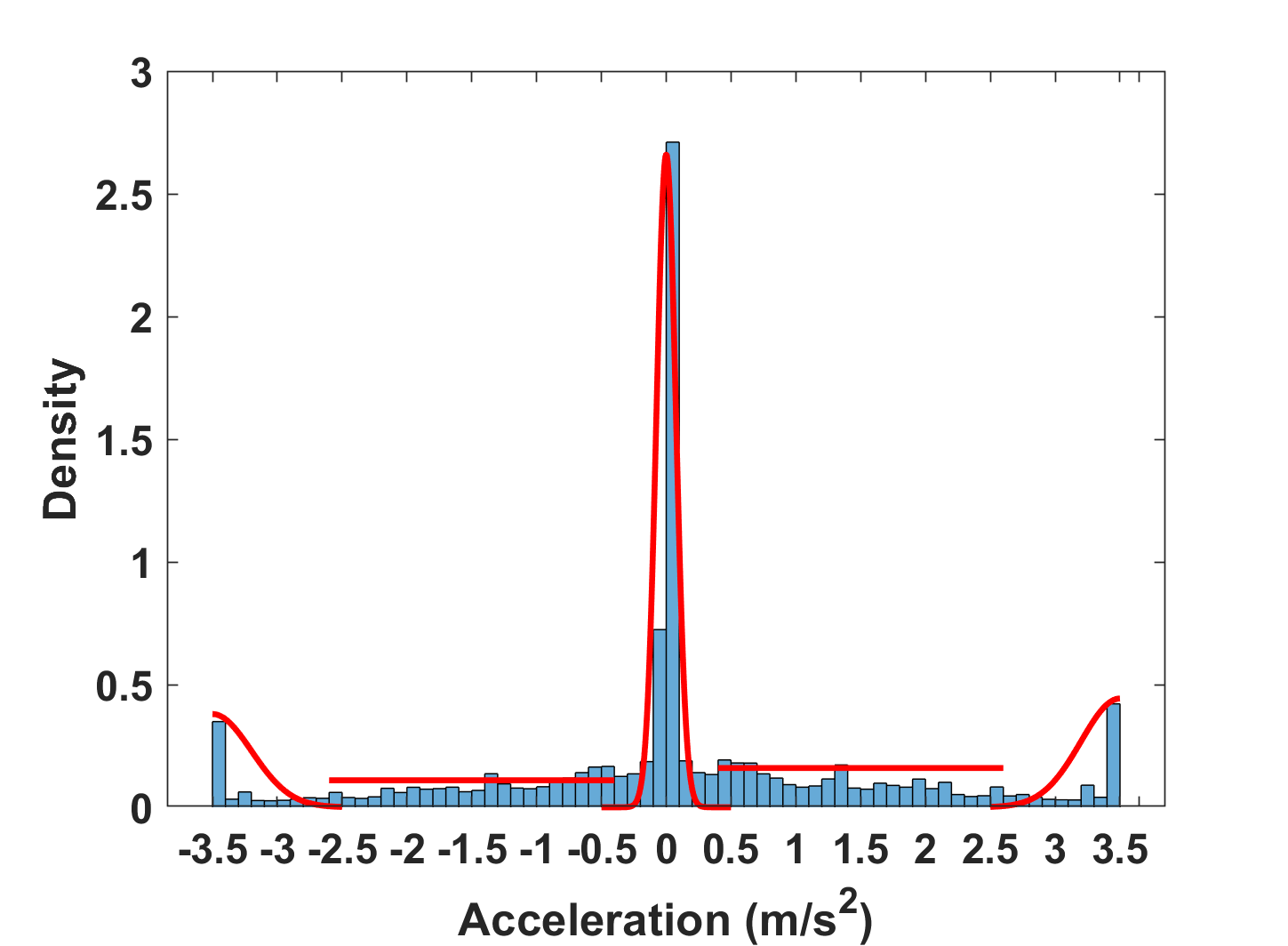}
	  \caption{Acceleration distribution is approximated with five different distributions: a normal distribution with 0 mean and 0.075 standard deviation, a uniform distribution between 0.5 m/s and 2.5 m/s, a uniform distribution between -0.5 m/s and -2.5 m/s, a half normal distribution with 3.5 mean and 0.3 standard deviation and a half normal distribution with -3.5 mean and 0.3 standard deviation. x-axis shows the accelerations in $\text{m/s}^2$ and y-axis presents the values of the probability density function.}
	\label{accandfit}
\end{figure}

Drivers have two action types: \textit{changing lane} and \textit{changing acceleration}. For lane change, two actions are defined: \textit{moving to the left} lane and \textit{moving to the right} lane. In order to determine \textit{acceleration changing} actions, the distribution of vehicle accelerations, obtained by processing the US101 data, is used. Figure \ref{accandfit} presents the acceleration distribution. On the figure, five regions are identified and approximated by known continuous distributions that are shown in red color and superimposed on the original figure. Based on this acceleration data analysis, the driver actions in terms of accelerations are defined as

\begin{enumerate}
\item \textit{Maintain}: acceleration is sampled from normal distribution with $\mu=0, \sigma = 0.0075 \text{m/s}^2$.
\item \textit{Accelerate}: acceleration is sampled from a uniform distribution between $0.5 \text{m/s}^2, 2.5 \text{m/s}^2$.
\item \textit{Decelerate}: acceleration is sampled from a uniform distribution between $-0.5 \text{m/s}^2, -2.5 \text{m/s}^2$.
\item \textit{Hard Accelerate}: acceleration is sampled from a inverse half normal distribution with $\mu=3.5 \text{m/s}^2, \sigma = 0.3 \text{m/s}^2$.
\item \textit{Hard Decelerate}: acceleration is sampled from a half normal distribution with $\mu=-3.5 \text{m/s}^2, \sigma = 0.3 \text{m/s}^2$.
\end{enumerate}

\begin{remark}
Distributions superimposed on the histogram in Fig. \ref{accandfit} are continuous. Therefore, all driver actions are sampled from continuous distributions. 
\end{remark}

\subsection{Physical Model of Vehicles}
As explained before, drivers can take lane changing or acceleration changing actions. It is assumed that changing the lane takes 1 second.

In Fig. \ref{fig:model1}, the variable $x$ is used to represent the longitudinal position and $y$ represents the lateral position. Similarly, $v_x$ and $v_y$ represent the longitudinal and lateral velocities, respectively. The equations of motion for the vehicles in the traffic are given by
\begin{equation}
x(t_0+t) = x(t_0) + v_x(t_0) + \frac{1}{2}a(t_0)(t-t_0)^2
\end{equation}
\begin{equation}
y(t_0+t) = y(t_0) + v_y(t_0)
\end{equation}
\begin{equation}
v_x(t_0+t) = v_x(t_0) + a(t_0)(t-t_0),
\end{equation}
where $t_0$ is the initial time-step and $a$ is the acceleration.

\subsection{Vehicle Placements}
At the beginning of the training and simulations, vehicles are randomly placed on a 600m endless circular road segment. Initial distances between vehicles are constrained to be larger than, or equal to, 11m, the upper limit of $close$. Initial velocities are selected to prevent impossible-to-handle cases at the beginning of the training or simulation: A driver who is in close proximity to the vehicle in front should be able to prevent a crash using the $hard~ decelerate$ action.

\subsection{Reward Function}
Rewards are collected based on a reward function, which represents the goals of the driver. In traffic, drivers try to avoid crashing and getting too close to other cars. Furthermore, minimizing the travel time with minimal effort is desired. In the reward function, a variable is defined for each of these goals and the weights are assigned to these variables to emphasize their relative importance. The reward function is defined as
\begin{equation}
R = w_1*c + w_2*s + w_3*d + w_4*e,
\end{equation}
where $w_i$ are the weights. The terms of the reward function are defined below.

$c$: Equals to -1 if a crash occurs and 0, otherwise. Penalizes the driver if an accident occurs. Getting out of the road boundaries is also considered as a crash.

$s$: Equals to the difference between the speed of the driver and the mean speed normalized by the maximum speed. Thus, higher velocities are rewarded to improve performance. The formula to calculate this term is
\begin{equation}
s= \frac{v(t) - \frac{v_{max}+v_{min}}{2}}{v_{max}}.
\end{equation} 

$d$: Equals to -1 if the distance to the car in front is $close$, 0 if the distance to the car in front is $nominal$, and 1, otherwise. Rewards keeping the headway large and penalizes small headways. 

$e$: Equals to 0 if the action of the driver is $maintain$, -0.25 if the action is $accelerate$ or $decelerate$, -0.5 if the action is $hard accelerate$ or $hard decelerate$ and -1 if the action is $move left$ or $move right$. Penalizes the effort consumed for the action taken.

\begin{remark}
The nominal velocity is selected as 12.29m/s (27.49mph), and the maximum allowed velocity is selected as two times of the nominal velocity, 24.59m/s (55mph), which is the speed limit at US101 for the selected road section.  Nominal velocity is not the desired velocity, in fact, as shown in (17), drivers take positive rewards as they approach to the maximum velocity, and penalized for velocities smaller than the nominal velocity.
\end{remark}

\section{Training and Simulation}
For the training of the driver policies, two separate reinforcement learning (RL) methods, Deep Q-Learning (DQN) and its the continuous version, c-DQN, are used together with the level-k reasoning approach. The advantages of these RL methods over each other are discussed in the previous section. The training environment is a 5-lane road, where a training episode is defined by a fixed number of simulation steps. When a crash occurs, the existing episode ends a new one is initialized.

During the training of a level-k driver, 125 level-(k-1) vehicles are placed on the road, together with the ego vehicle. This makes the density of the cars approximately equal to the average car density in the US101 data \cite{US101}. The number of cars is decreased to 100 at the end of the $1300_{th}$ episode and increased to 125 again at the end of the $3800_{th}$  episode, to increase the number of states that the drivers are exposed to during training.

Both DQN and c-DQN algorithms are implemented using the Phyton library Keras \cite{chollet2015keras}, together with the stochastic optimizer Adam \cite{kingma2014adam}. The initial learning rate is selected as 0.005, the discount factor $\gamma$ is selected as 0.975 and the memory capacity $N$ is selected as 2000. 

\subsection{Level-0 Policy}
The non-strategic level-0 policy must be determined first, before obtaining other levels. A level-0 policy can be defined by using several different approaches. For instance, a uniformly random selection of actions can be defined as level-0 policy \cite{shapiro2014level}. In earlier studies, where approaches similar to the one proposed in this paper, level-0 policies are set as a persisting single action regardless of the state being observed \cite{musavi2016unmanned}, \cite{musavi2016game}, \cite{yildiz2013predicting}, or as a conditional logic based on experience \cite{backhaus2013cyber}. The level-0 policy used in this study is defined as
\begin{enumerate}
\item $hard\;decelerate$ if the car in front is $close$ and $approaching$;
\item $decelerate$ if the car in front is $close$ and $stable$ or $nominal$ and $approaching$;
\item $accelerate$ if the car in front is $nominal $ and $moving away$ or $far$ and
\item $maintain$ otherwise.
\end{enumerate}

\subsection{Computing Environment}
Specifications of the computer used in training are given as
\begin{itemize}
\item Processor: Intel® Core™ i7 8750H
\item RAM: 32GB DDR4 2133 MHz
\item GPU: NVIDIA® GeForce® RTX 2070 8GB
\end{itemize}

With this computer, for 5000 episodes, each of which contains 100 steps, training of each level took approximately 20 hours both with DQN and c-DQN.

\subsection{Training Performance}
\begin{figure}[h!]
  \includegraphics[width=\linewidth]{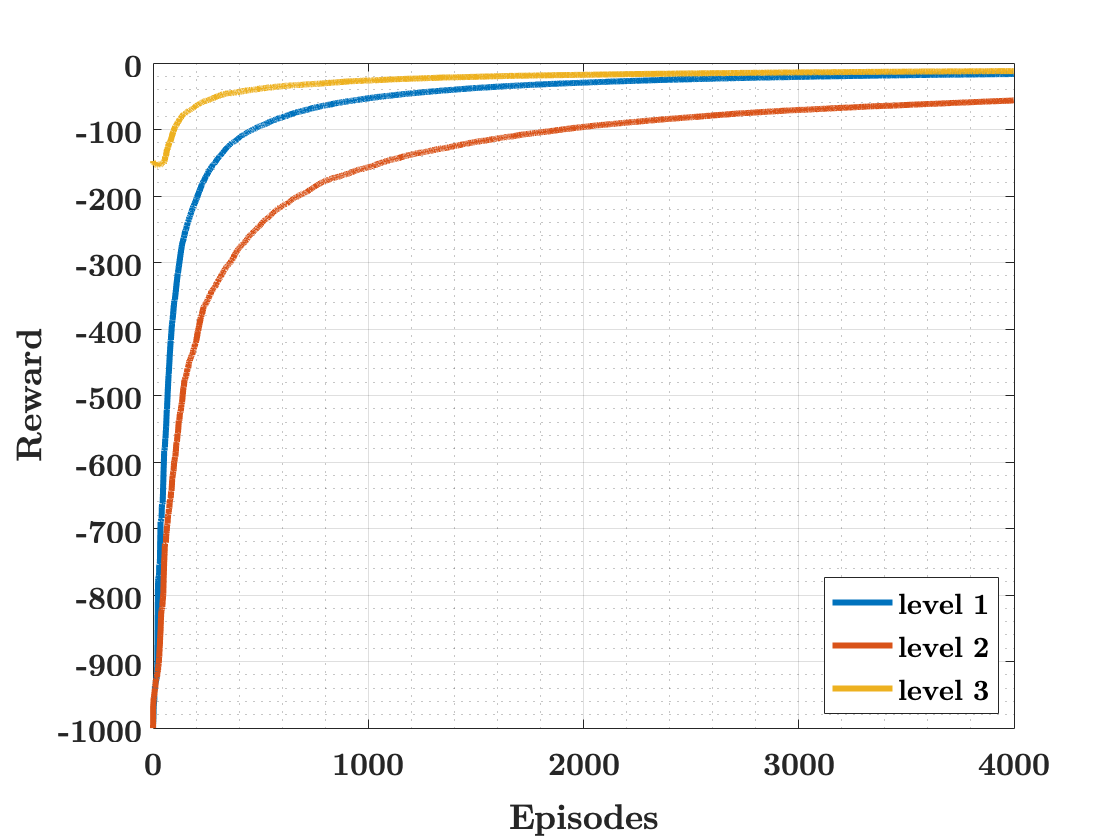}
  \caption{Average rewards during level-1, level-2 and level-3 policy training for Deep Q-Learning.}
  \label{fig:rew_dqn}
\end{figure}

\begin{figure}[h!]
  \includegraphics[width=\linewidth]{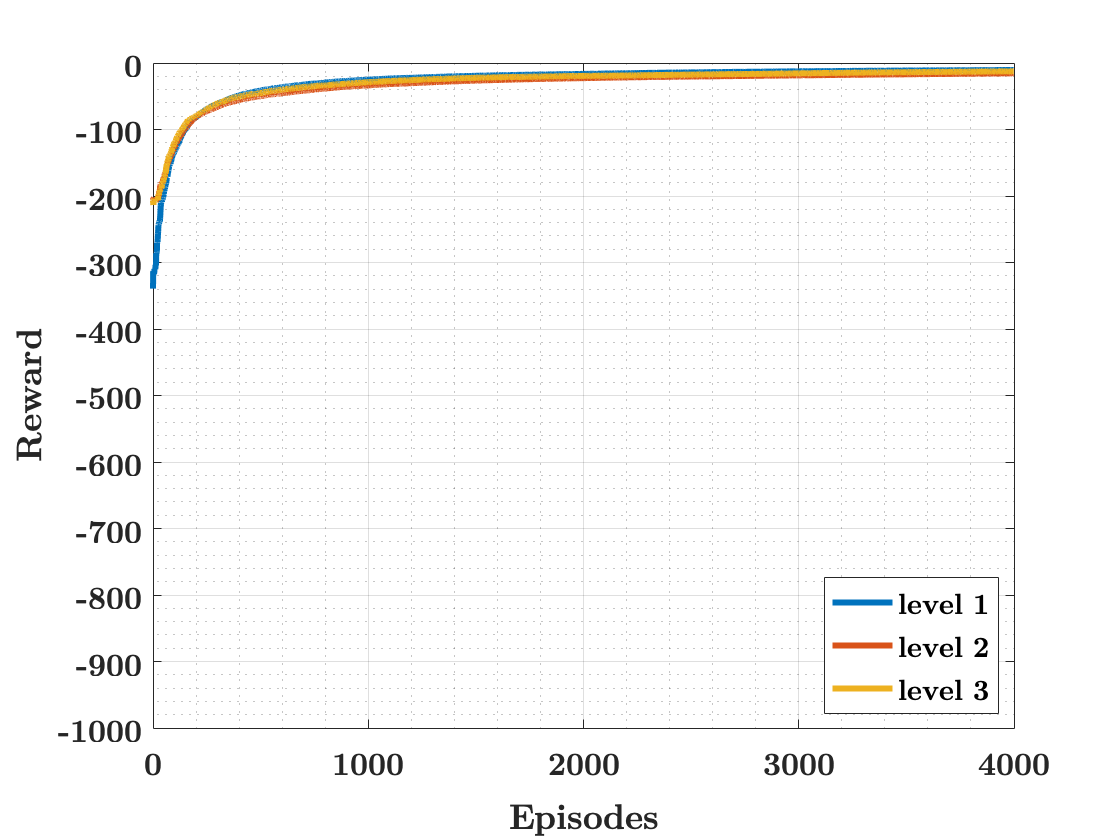}
  \caption{Average rewards during level-1, level-2 and level-3 policy training for continuous Deep Q-Learning.}
  \label{fig:rew_cdqn}
\end{figure}

Fig. \ref{fig:rew_dqn} and Fig. \ref{fig:rew_cdqn}, show the evolution of the average rewards during training, for DQN and c-DQN methods, respectively. The rewards monotonically increase and eventually converge for both of the methods. c-DQN rewards tend to have a more uniform structure within different levels compared to the regular DQN. The main reason of this behavior seems to be that c-DQN learns and convergences faster and therefore reward curves look more uniform when plotted with the same scale.

\subsection{Simulation Performance}
\begin{figure}[h!]
  \includegraphics[width=\linewidth]{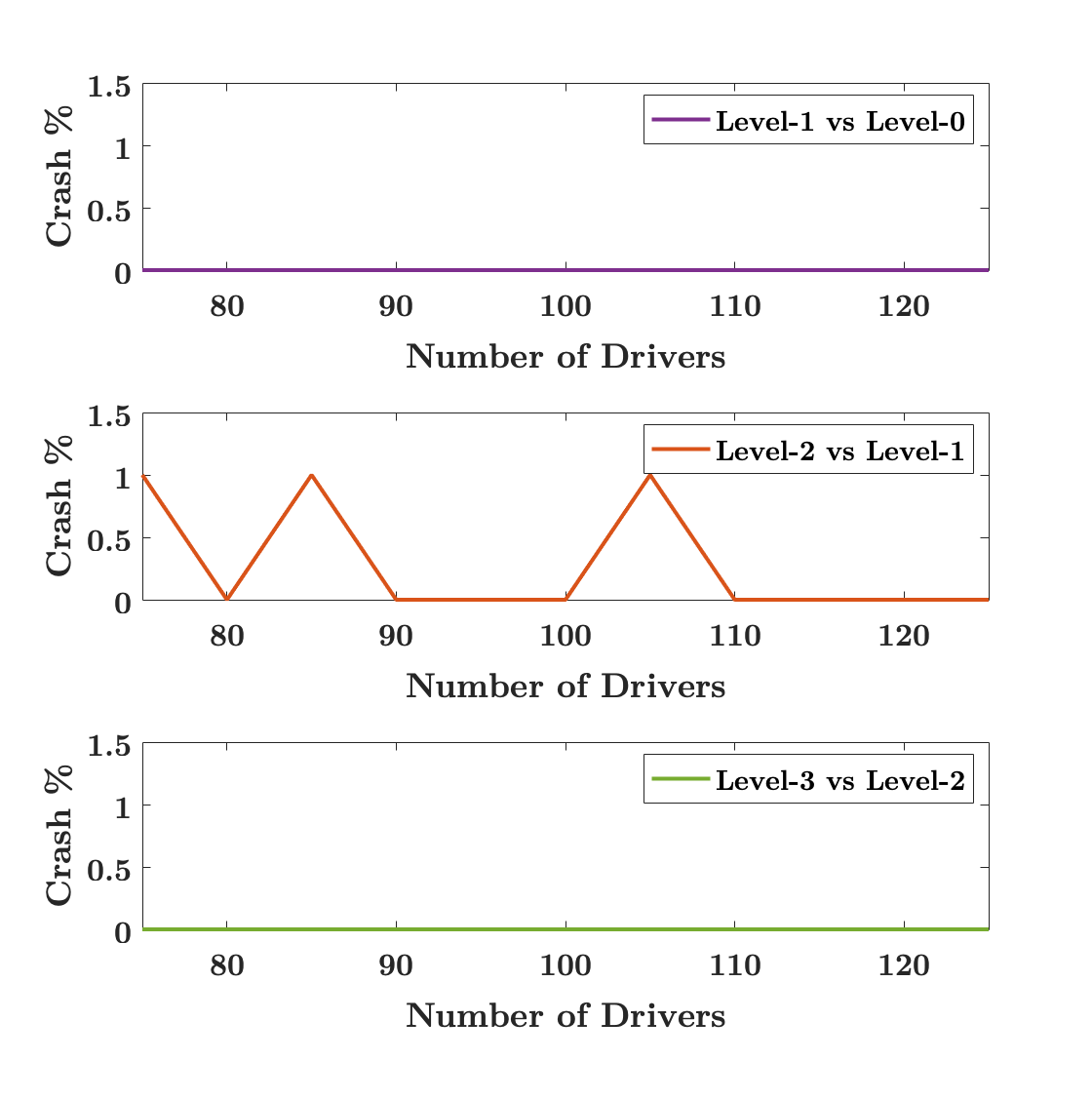}
  \caption{Crash rates for level-k vs level-(k-1) scenarios for different number of cars on a circular 600m road for policies trained with DQN.}
  \label{crashdqn}
\end{figure}

\begin{figure}[h!]
  \includegraphics[width=\linewidth]{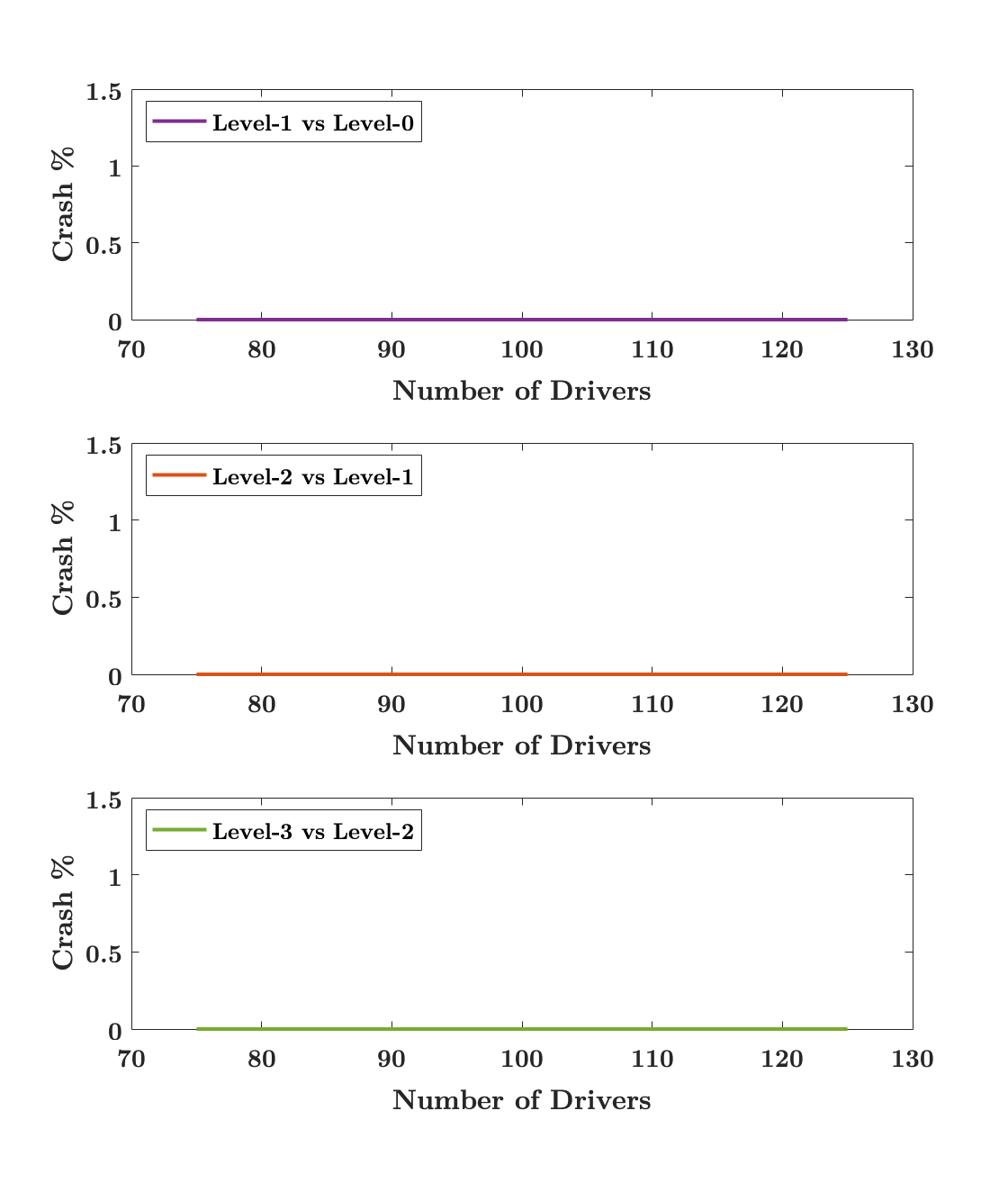}
  \caption{Crash rates for level-k vs level-(k-1) scenarios for different number of cars on a circular 600m road for policies trained with c-DQN.}
  \label{crashcdqn}
\end{figure}

The following scenarios are simulated
\begin{itemize}
\item Level-1 driver is placed on a traffic environment consisting $n_{d} − 1$ level-0 drivers on a 600m road segment 
\item Level-2 driver is placed on a traffic environment consisting $n_{d} − 1$ level-1 drivers on a 600m road segment 
\item Level-3 driver is placed on a traffic environment consisting $n_{d} − 1$ level-2 drivers on a 600m road segment,
\end{itemize}
where $n_d$ corresponds to the total number of drivers on the road. Simulations are performed for $n_d = 75,80,85,90,95, 100, 105, 110, 115, 120$ and  $125$, for each scenario. In all of the above scenarios, simulations are run for 100 episodes, each covering a 100s simulation. Simulation results, in terms of crash rates, for DQN and c-DQN are presented in Fig. \ref{crashdqn} and \ref{crashcdqn}, respectively. 

When compared with previous studies, \cite{li2017game} and \cite{albaba2019stochastic}, in terms of crash rates, policies proposed in this paper show more realistic driving behavior, since average crash rate is 2 per million miles driven nationally \cite{national2017crash}.

\section{Validation with Traffic Data}
In order to compare the proposed policies, i.e. driver models, with the policies obtained by processing the real traffic data, Kolmogorov Smirnov Goodness of Fit Test (K-S Test) is employed. This is one of the most commonly used non-parametric goodness of fit test [53]. Since the policies consist of discrete probability distributions, K-S Test for Discontinuous Distributions \cite{conover1972kolmogorov} is used. The test is explained briefly in the following section and a more detailed description can be found in \cite{conover1972kolmogorov}.

\subsection{Kolmogorov-Smirnov Test for Discontinious Distributions}
For an unknown discrete probability distribution function (pdf) $F(x)$ and a hypothesized pdf $H(x)$, the null hypothesis of the K-S Test is
\begin{equation}
H_0: F(x) = H(x) \text{ for all x.}
\end{equation}
To test the null hypothesis, first, empirical cumulative pdf of observed data, $S_n(x),$ and hypothesized cumulative pdf, $H_c(x),$ are calculated. Secondly, the \text{test statistic}s, which are measures of the difference between $S_n (x)$ and $H_c(x)$, are calculated as
\begin{equation}
D = sup_x |H_c(x) - S_n(x)|
\end{equation}
\begin{equation}
D ^-= sup_x (H_c(x) - S_n(x))
\end{equation}
\begin{equation}
D^+ = sup_x (S_n(x) - H_c(x)),
\end{equation}
where $D$ is the two sided, and $D^-$ and $D^+$ are one sided test statistics. Thirdly, \textit{critical level}s of $D^-$ and $D^+$, $P(D^- \leq d^-)$ and $P(D^+ \leq d^+)$, are calculated using Algorithm 5-6, respectively, where $n$ denotes the sample size, and $d^-$ and $d^+$ denote the observed values of $D^-$ and $D^+$. 

\begin{algorithm}[H]
	\caption{Calculation of the critical value of $D^-$}
	\begin{algorithmic}[1]
		\FOR {i = 1 to $n*(1-d^-)$} 
		\STATE On the graph of $H_c(x)$, draw a horizontal line with ordinate $d^- + i/n$
		\IF {this line intersects with the graph of $H_c(x)$ at a jump (a discontinuity),}
		\STATE Set $c_i=1-H_c(x)$. If the intersection occurs exactly at the left limit of the discontinuity, use the left limit value for $H_c(x)$. Otherwise use the right limit.
		\ELSE
		\STATE Set $c_i = 1-d^--i/n$
		\ENDIF
		\ENDFOR
		\STATE Set $b_0 = 1$
		\FOR {i = 1 to $n*(1-d^-)$}
		\IF {$c_i$ $ > $ 0}
		\STATE Set $b_i = 1 - \sum_{j=0}^{i-1}C(i,j)*c_j^{i-j}*b_i$
		\ENDIF
		\ENDFOR
		\STATE Calculate the critical level as:
		\STATE $P(D^- \geq d^-) = \sum_{i=0}^{n*(1-d^-)}C(n,i)*c_i^{n-i}*b_i$
	\end{algorithmic}
\end{algorithm}

\begin{algorithm}[H]
	\caption{Calculation of the critical value of $D^+$}
	\begin{algorithmic}[1]
		\FOR {i = 1 to $n*(1-d^+)$} 
		\STATE On the graph of $H_c(x)$, draw a horizontal line with ordinate $1-d^+- i/n$
		\IF {This line intersects with the graph of $H_c(x)$ at a jump (a discontinuity),}
		\STATE Set $c_i=1-H_c(x)$. If the intersection occurs exactly at the right limit of the discontinuity, use the right limit value for $H_c(x)$. Otherwise use the left limit.
		\ELSE
		\STATE Set $f_i = 1-d^+-i/n$
		\ENDIF
		\ENDFOR
		\STATE Set $e_0 = 1$
		\FOR {i = 1 to $n*(1-d^+)$}
		\IF {$f_i$ $>$ 0}
		\STATE Set $e_i = 1 - \sum_{j=0}^{i-1}C(i,j)*f_j^{i-j}*e_i$
		\ENDIF
		\ENDFOR
		\STATE Calculate the critical level as:
		\STATE $P(D^+ \geq d^+) = \sum_{i=0}^{n*(1-d^+)} C(n,i)*f_i^{n-i}*e_i$
	\end{algorithmic}
\end{algorithm}

Finally, the critical value for the two-sided test statistic is determined as
\begin{equation}
P(D \geq d) \doteq P(D^+ \geq d) + P(D^- \geq d),
\end{equation}
where d is the observed value of D. It is noted that this critical value describes the percentage of data samples, whose test statistics are larger than or equal to $d$, given that the null hypotheses is true. Thus, the probability of an observation (data point) being sampled from the hypothesized model, $H(x)$, or, equivalently, the probability of the null hypothesis being true, increases with the increase in the critical value. The null hypothesis is rejected if the critical value is smaller than a certain threshold called the \textit{significance value} $\alpha$, which is selected as 0.05 in this work.

\begin{remark}
When the null hypothesis can not be rejected, it means that there are not enough data-based evidence that the investigated model is not representative of the real data. This leads to retaining the null hypothesis. Therefore, in the rest of the paper, we call a data-modal comparison ``successful" when the null hypothesis is not rejected. 
\end{remark}

\subsection{Comparing game theoretical models with traffic data}
Obtaining game theoretical (GT) policies, which are stochastic maps from observations to actions, is explained in previous sections. To obtain the data-based driver policies, the traffic data \cite{I80} and \cite{US101} are processed and for each vehicle, action probability distributions over all visited states are generated. The probabilities are calculated by the frequencies of the actions taken by the drivers for a given state. Action probabilities that are lower than 0.01 are replaced with 0.01 with re-normalizations in order to eliminate close-to-zero probabilities, for both the GT policies and the ones obtained from the data.

GT and the data-based policies are compared for each individual driver: For a given vehicle, whose states and actions are captured in the traffic data, first, the vehicle driver's frequency of actions for each visited state are calculated. These frequencies are converted to probability distributions over actions, for each state. The probability distributions are called the policies of the driver. Once the driver policies are determined from the data, these policies are compared with the GT policies, using the K-S test, for each state. Finally, \textit{success rate} of the GT policies, for the individual driver being investigated, is defined as the ratio of the states whose corresponding policies are successfully modeled by the GT policies over all the visited states. For example, the result may state that ``70\% of the states visited by Driver-1 are successfully modeled by the GT policies, therefore the success rate is 70\%". The process of comparison for each driver is given in Algorithm 7, where $nstate$ defines the number of states that are visited by the driver, $n^i_{visit-driver}$ defines the number of times the state $i$ is visited by the driver. Furthermore, $n_{comparisons}$ and $n_{success}$ are the total number of states whose policies are compared and the number of successful comparisons, respectively. Since K-S test works best for large sample sizes, the comparisons are conducted for states that are visited more than a certain number, which we call $n_{limit}$.

\begin{algorithm}[H]
	\caption{Comparing GT models with traffic data, for an individual driver}
	\label{kolmogorov}
	\begin{algorithmic}[1]
		\FOR {i = 1 to $nstate$} 
			\IF{$n^i_{visit-driver}  \geq n_{limit}$}
			
				\STATE $n_{comparisons}+=1$
				\STATE Set $p_i$ to the GT policy (pdf).
				\STATE Set $k_i$ to the data-based policy (pdf).
					\STATE Set $H_c$ to the cumulative pdf obtained from $p_i$.
					\STATE Set $S_n$ to the cumulative pdf obtained from $k_i$.
					\STATE Test the null hypothesis (18) using K-S test.
				\IF{Null hypothesis is not rejected}
					\STATE $n_{success} += 1$
				\ENDIF
			\ENDIF
		\ENDFOR
		\STATE Percentage of successfully modeled states (for this specific driver) = $100\frac{n_{success}}{n_{comparisons}}$
		
	\end{algorithmic}
\end{algorithm}

Data-based policies of each individual driver are compared with the proposed GT policies using Algorithm 7 and the \textit{success rate}s for each driver are found and plotted. 

It is noted that the proposed GT models are pdfs over the action space defined in Chapter III. To compare the performance of these models with an alternative model, the alternative model should have the same stochastic map structure where the pdfs are given over the same action space. If this requirement is not satisfied, i.e. the alternative model is not stochastic or does not have the same action space, then it becomes unclear how to conduct a systematic comparison. Furthermore, in the case that the domain of the stochastic map (action space) is not the same, then the traffic data needs to be reprocessed to obtain policies that have the same structure of the alternative model, which is a nontrivial task. A commonly used method in these circumstances is creating a benchmark model which is used a minimum performance threshold for the tested model. In this study, we used a model that has uniform pdf over the action space as the benchmark and the \textit{success rate}s (see Algorithm 7) of the benchmark model are also provided for comparison purposes. 

\subsection{Results}

For the validation of the proposed game theoretical (GT) driver models, two different sets of traffic data, obtained from the highways US101 \cite{US101} and I80 \cite{I80}, are used.

The following definitions are employed when reporting the validation results.
\medskip

\begin{definition}
Given two discrete probability distribution functions (pdf) p and q, the \textit{Mean Absolute Error} (MAE) between p and q is defined as
\begin{equation}
MAE=1/n \sum_{i=1}^n |p(x_i)-q(x_i)|,
\end{equation}
where $x_i$s are random variables.
\end{definition}

\begin{definition}
$aMAE$ is the average of the $MAE_js$ between the GT policies and the data-based policies, for which the null hypothesis is not rejected. Therefore, $aMAE$ is calculated as
\begin{equation}
aMAE=1/M \sum_{j=1}^m MAE_j,
\end{equation}
where M is the number of comparison for which the null hypotheses is not rejected. 
\end{definition}

\begin{definition}
$rMAE$ is the average of the $MAE_ks$ between the GT policies and the data-based policies, for which the null hypothesis is rejected. Therefore, $rMAE$ is calculated as
\begin{equation}
rMAE=1/K \sum_{k=1}^m MAE_k,
\end{equation}
where K is the number of comparison for which the null hypotheses is rejected. 
\end{definition}

Model vs data comparisons are made for two different $n_{limit}$ values, specifically for $n_{limit} = 3$ and $n_{limit} = 5$. As explained earlier, $n_{limit}$ is the minimum number of state visits in the traffic data, for the corresponding policy to be considered in the K-S test. When a state is visited a small number of times, the probability of the resulting driver policy for this state being sampled from a Uniform action-probability-Distribution (UD) increases. This means that the policy for the given state does not have a ``structure" that is distinguishable from a UD. It is observed that the minimum number of state visits is approximately equal to 3 for the K-S test to acknowledge that the policy is sampled from a non-uniform distribution, with a significance value of 0.05. Therefore we report the results for $n_limit=3$. Moreover, we also report the results for $n_{limit}=5$ to demonstrate the effect of this variable on the test outcomes. Finally, we state the $RL_{method}$ in the results, which is the reinforcement learning method, either DQN or c-DQN, used in the tests. 

\subsubsection{Model validation using US-101 Data}
In this section, we give comparison results between the policies obtained by processing the raw US-101 Data and the GT policies. The data are collected between 7.50-8.05 AM and consists of 2168 different drivers \cite{US101}.

\medskip
\noindent \textit{a)$RL_{method}:$DQN, $n_{limit}=3$}
\smallskip

For this model-data comparison, $aMAE = 0.50$ and $rMAE = 1.55$.

\begin{figure*}
     \centering
     \begin{subfigure}[t]{0.49\textwidth}
        \centering
		\includegraphics[width=\textwidth]{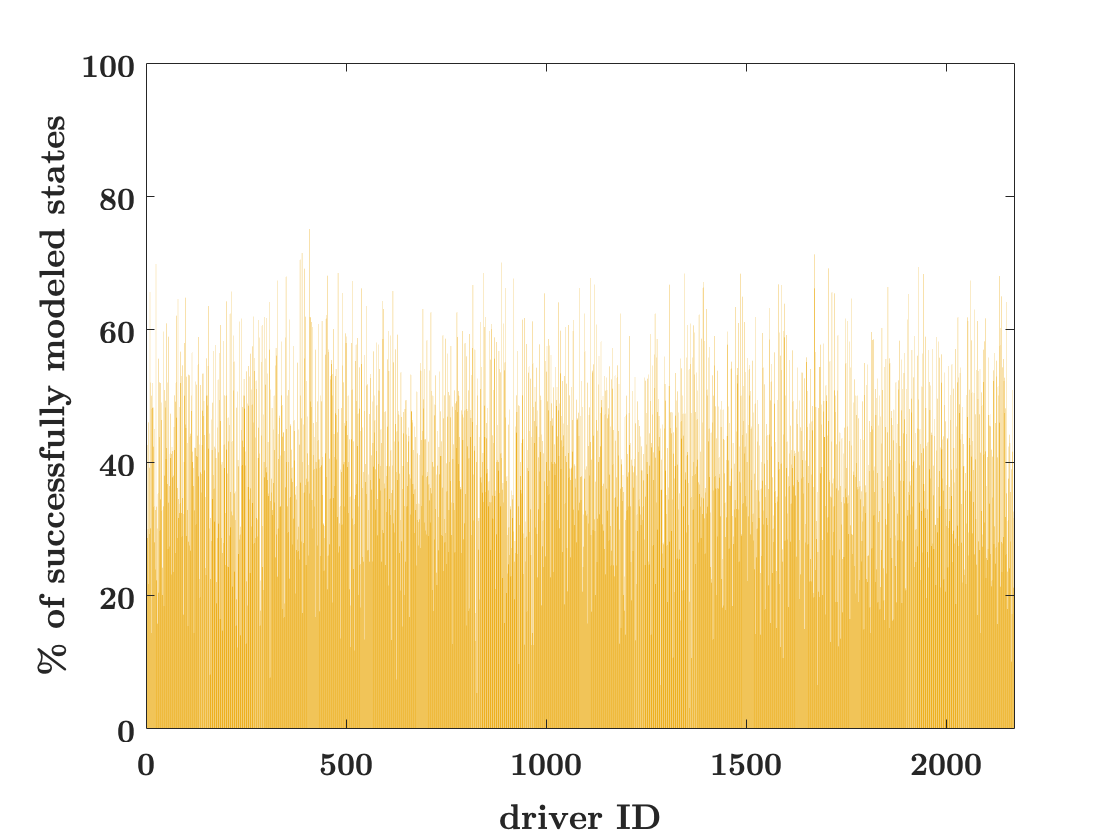}
		\caption{Percentages of successfully modeled states by the GT policies obtained through DQN, for each driver. Each vertical line belongs to an individual driver.}
		\label{fig:policy3d}
     \end{subfigure}
     \hfill
     \begin{subfigure}[t]{0.49\textwidth}
        \centering
		\includegraphics[width=\textwidth]{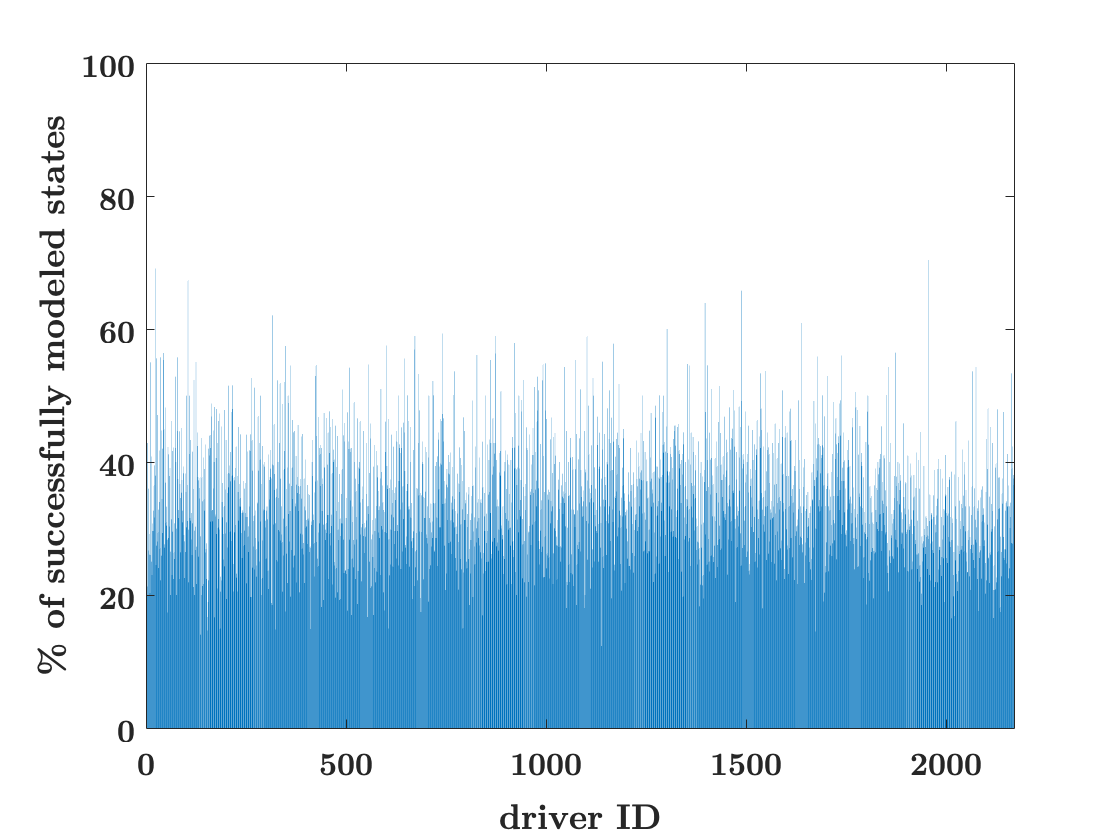}
		\caption{Percentages of successfully modeled states by the UD policy, for each driver. Each vertical line belongs to an individual driver..}
		\label{fig:dumb3d}
     \end{subfigure}
     \hfill
     \begin{subfigure}[t]{0.49\textwidth}
        \centering
		\includegraphics[width=\textwidth]{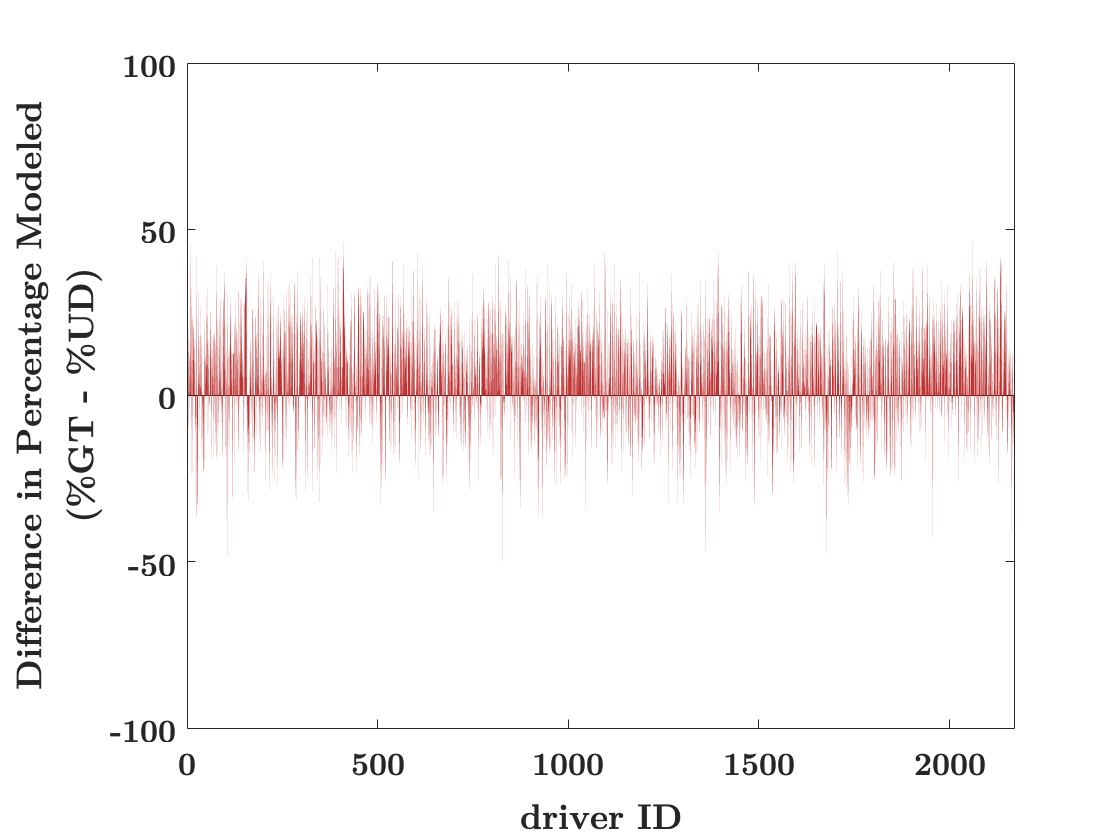}
		\caption{Differences in the percentages of the successfully modeled states of each driver, between the DQN-based GT policies and the UD policy.}
		\label{fig:dif3d}
     \end{subfigure}
     \hfill
     \begin{subfigure}[t]{0.49\textwidth}
        \centering
		\includegraphics[width=\textwidth]{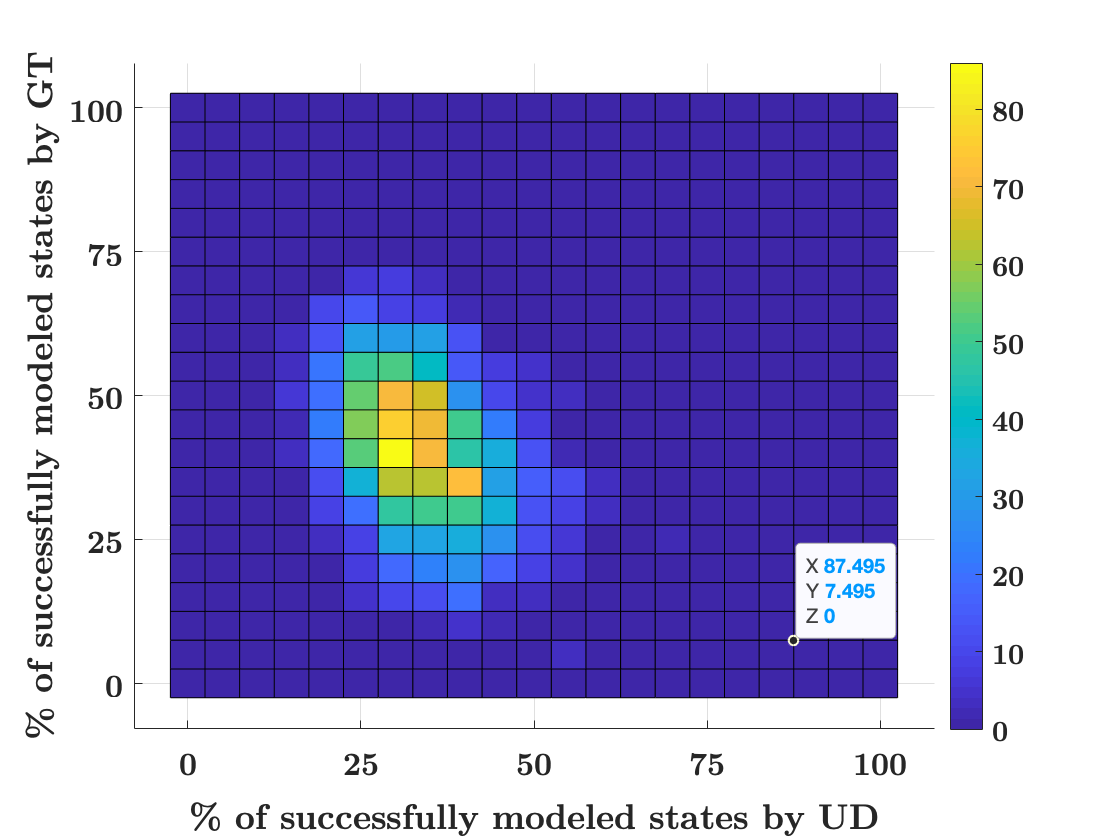}
		\caption{Color map showing the number of drivers whose x\% of the visited states are successfully modeled by the UD policy and y\% by the DQN-based GT policy. x and y percentages are given in the horizontal and vertical axes, respectively. }
		\label{fig:color3d}
     \end{subfigure}
        \caption{Comparison results for $n_{limit} = 3$ and $RL_{method}=DQN$ (US101)}
        \label{fig:res3d}
\end{figure*}

Fig. \ref{fig:policy3d} and \ref{fig:dumb3d} shows the performances of the proposed GT policies and the uniform distribution (UD) policy in terms of modeling human driver behaviors. In these figures, the x-axis shows driver IDs, which start from 1 and ends at 2168; and the y-axis shows the percentages of the successfully modeled states, for each driver. The differences between the percentages of successfully modeled states, for each driver, by the GT policies and the UD policy are presented in \ref{fig:dif3d}. Overall, Figs. \ref{fig:policy3d}-c show that the performance (success rate) of the GT policies are better than the UD policy in general. However, the difference is not large. Fig. \ref{fig:color3d} shows the difference between the performances of the GT policies and the UD policy using a different visualization method: In the figure, the x and y axes (horizontal and vertical) show the percentages of the successfully modeled states by the UD and GT policies, respectively. The colors on the figure represent the number of drivers. For example, the figure shows that there are around 75 drivers, whose 50\% of the states' policies could be successfully modeled by the GT policy while only 30\% could be modeled by the UD policy. The colored cluster being above the x=y line show that GT performs better than the UD policy, in general.

\medskip
\noindent \textit{b)$RL_{method}:$DQN, $n_{limit}=5$}
\smallskip

For this model-data comparison, $aMAE = 0.42$ and $rMAE = 1.53$.

\begin{figure*}
     \centering
     \begin{subfigure}[t]{0.49\textwidth}
        \centering
		\includegraphics[width=\textwidth]{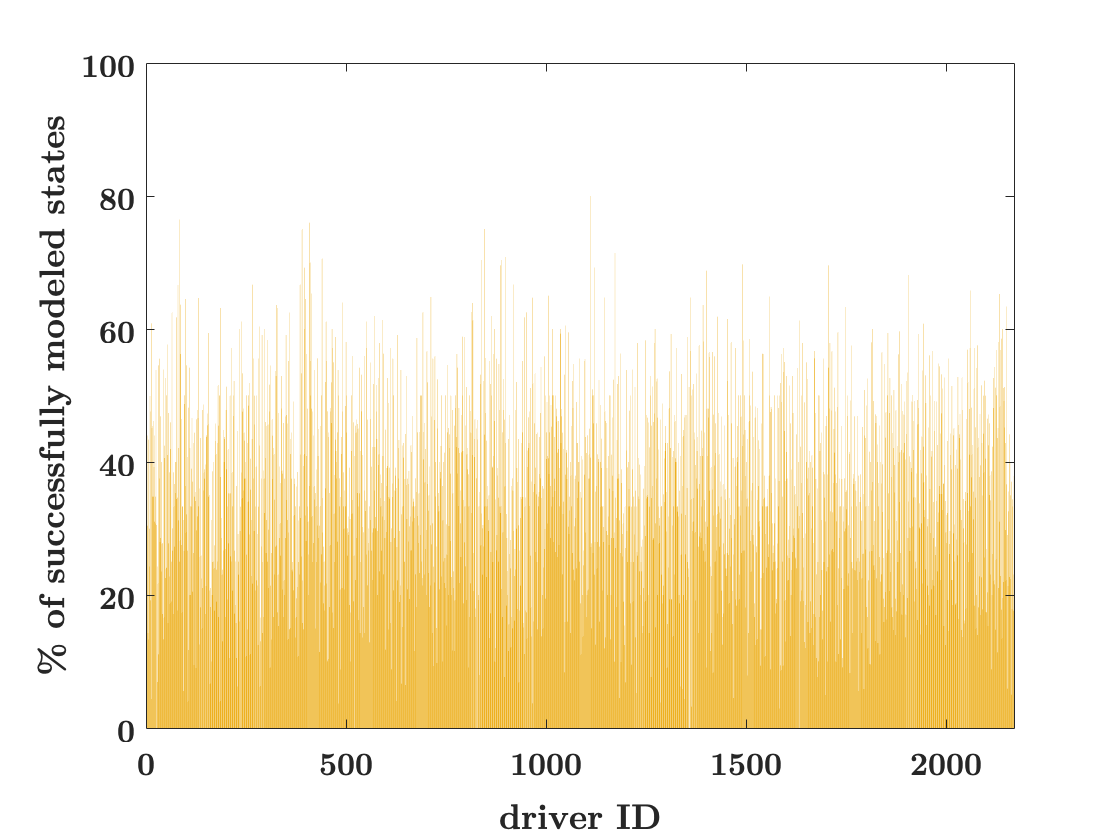}
		\caption{Percentages of successfully modeled states by the GT policies obtained through DQN, for each driver. Each vertical line belongs to an individual driver.}
		\label{fig:policy5d}
     \end{subfigure}
     \hfill
     \begin{subfigure}[t]{0.49\textwidth}
        \centering
		\includegraphics[width=\textwidth]{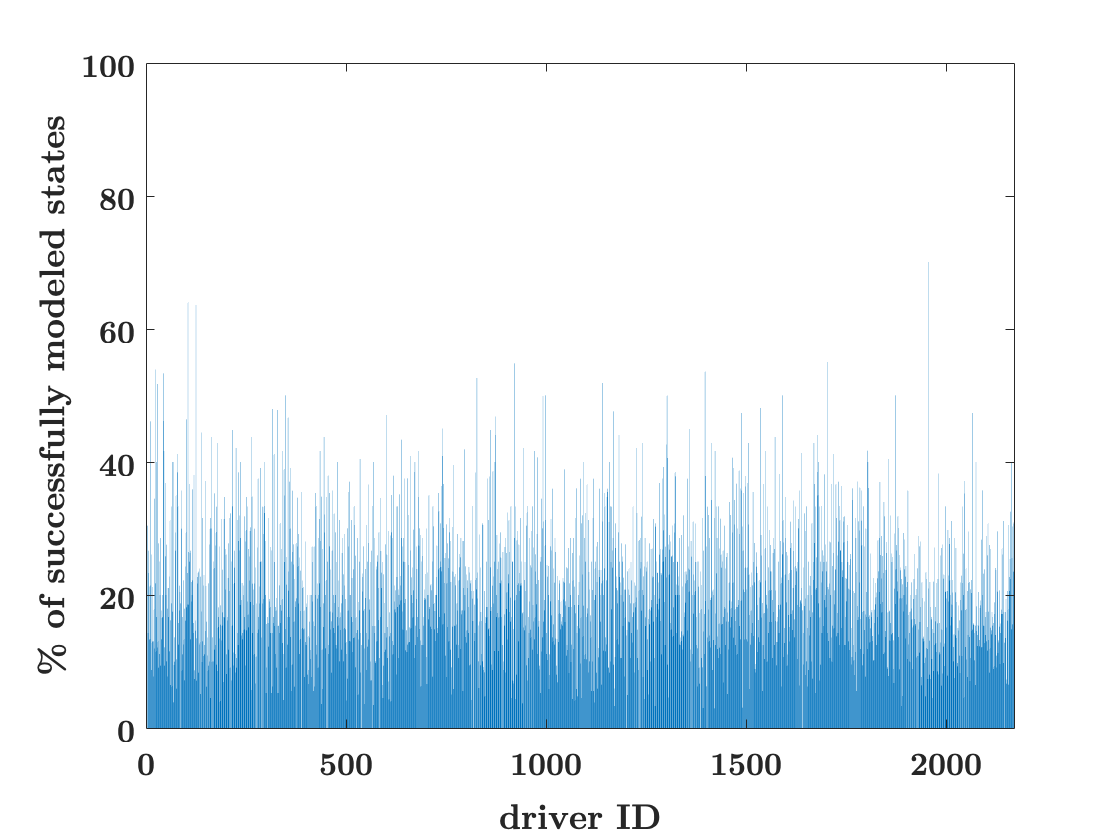}
		\caption{Percentages of successfully modeled states by the UD policy, for each driver. Each vertical line belongs to an individual driver..}
		\label{fig:dumb5d}
     \end{subfigure}
     \hfill
     \begin{subfigure}[t]{0.49\textwidth}
        \centering
		\includegraphics[width=\textwidth]{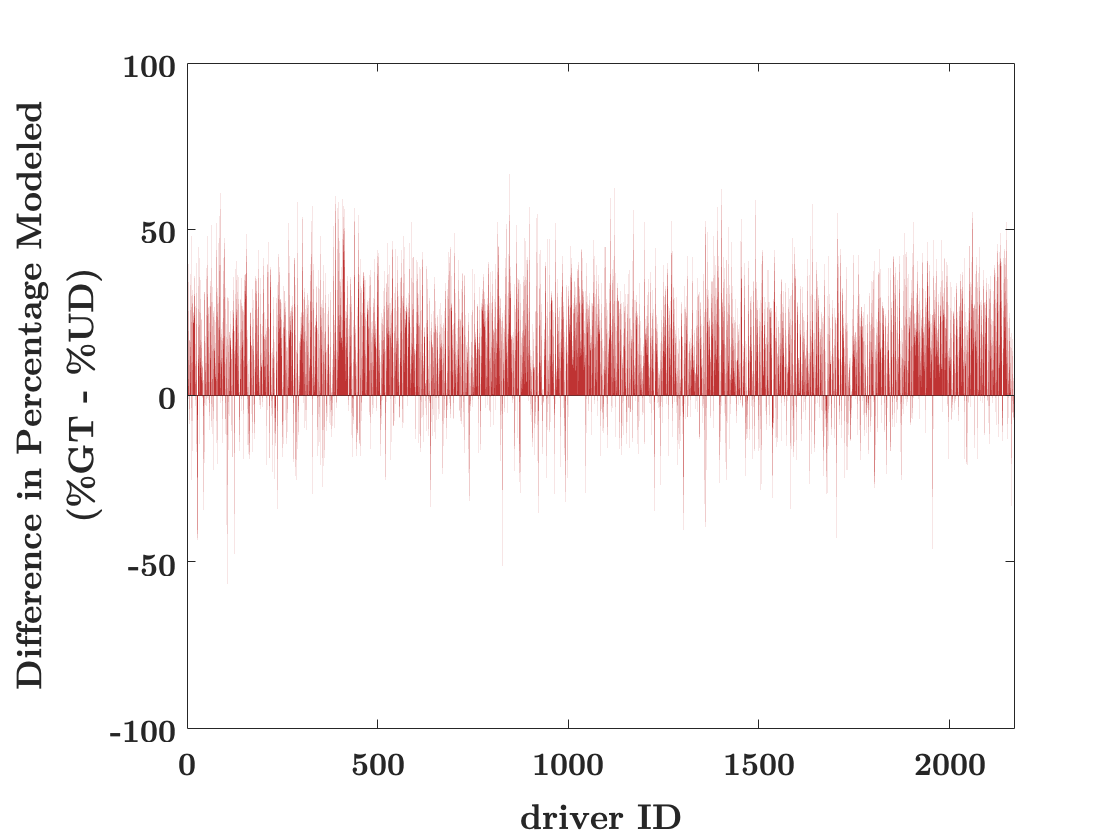}
		\caption{Differences in the percentages of the successfully modeled states of each driver, between the DQN-based GT policies and the UD policy.}
		\label{fig:dif5d}
     \end{subfigure}
     \hfill
     \begin{subfigure}[t]{0.49\textwidth}
        \centering
		\includegraphics[width=\textwidth]{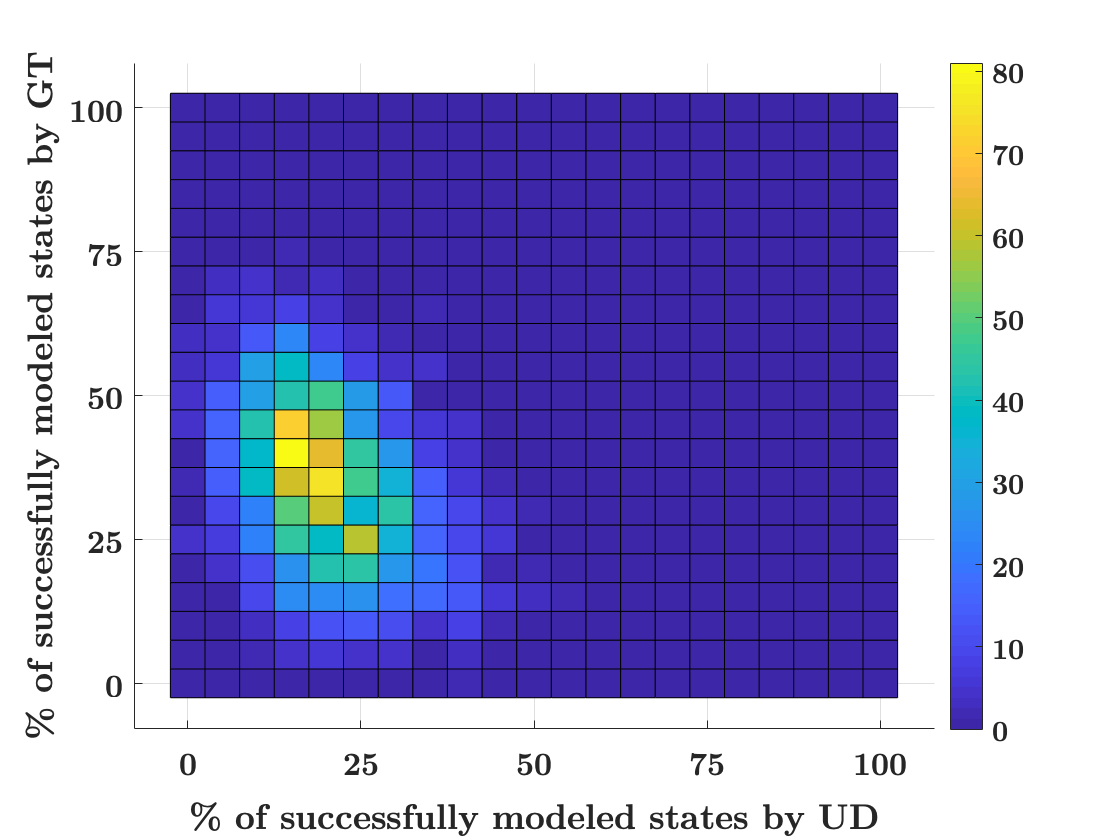}
		\caption{Color map showing the number of drivers whose x\% of the visited states are successfully modeled by the UD policy and y\% by the DQN-based GT policy. x and y percentages are given in the horizontal and vertical axes, respectively. }
		\label{fig:color5d}
     \end{subfigure}
        \caption{Comparison results for $n_{limit} = 5$ and $RL_{method}=DQN$ (US101)}
        \label{fig:res5d}
\end{figure*}

Performance of the GT and UD policies are shown in Figs. \ref{fig:policy5d} and \ref{fig:dumb5d}, respectively. The difference between these policies, in terms of successfully modeled state percentages, are given in Fig \ref{fig:dif5d}. When compared with Fig. \ref{fig:dif3d}, Fig. \ref{fig:dif5d} shows that the positive performance difference between the GT policies and the UD policy increases with the increase in $n_{limit}$ value. The main reason behind this is that with the increase of $n_{limit}$, the K-S test power increases and therefore the test can make better decisions in terms of determining whether or not the observed data is sampled from the hypothesized probability distribution function (pdf). The color map created earlier and presented in Fig. \ref{fig:color3d} is also created here but this time for $n_{limit}=5$, which is given in Fig. \ref{fig:color5d}. Compared to Fig. \ref{fig:color3d}, it is seen that the color cluster's distance from the x=y line is increased, corresponding to increased performance improvement of the GT policies over UD policy.

\medskip
\noindent \textit{c) $RL_{method}:$c-DQN, $n_{limit}=3$}
\smallskip

$aMAE = 0.73$ and $rMAE = 1.56$, for this model-data comparison.

\begin{figure*}
     \centering
     \begin{subfigure}[t]{0.49\textwidth}
        \centering
		\includegraphics[width=\textwidth]{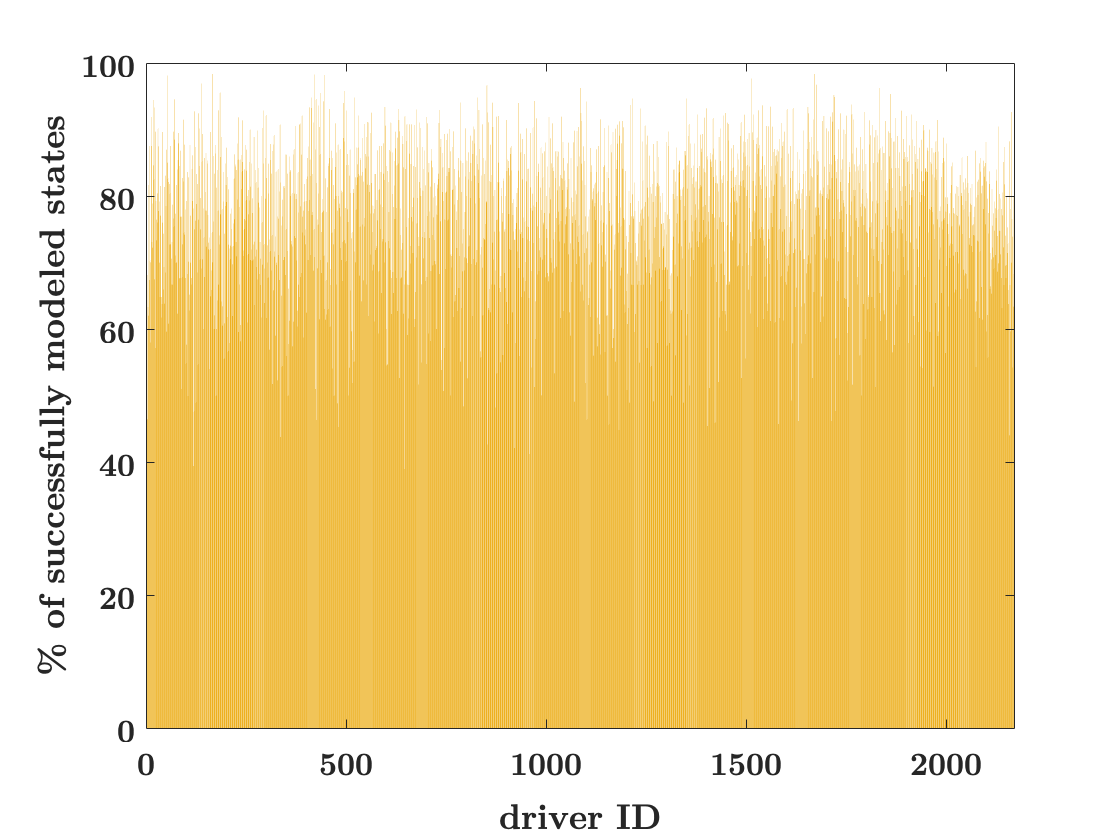}
		\caption{Percentages of successfully modeled states by the GT policies obtained through c-DQN, for each driver. Each vertical line belongs to an individual driver. }
		\label{fig:policy3c}
     \end{subfigure}
     \hfill
     \begin{subfigure}[t]{0.49\textwidth}
        \centering
		\includegraphics[width=\textwidth]{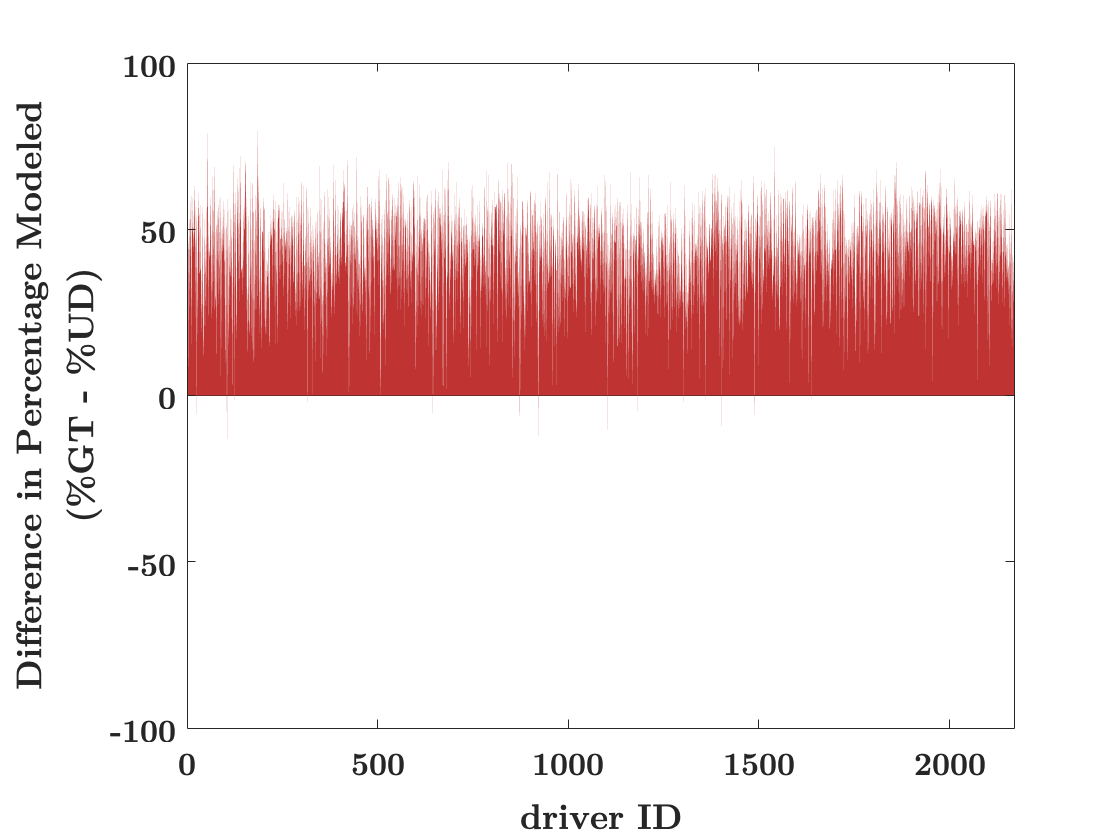}
		\caption{Differences in the percentages of the successfully modeled states of each driver, between the c-DQN-based GT policies and the UD policy.}
		\label{fig:dif3c}
     \end{subfigure}
     \hfill
     \begin{subfigure}[t]{0.49\textwidth}
        \centering
		\includegraphics[width=\textwidth]{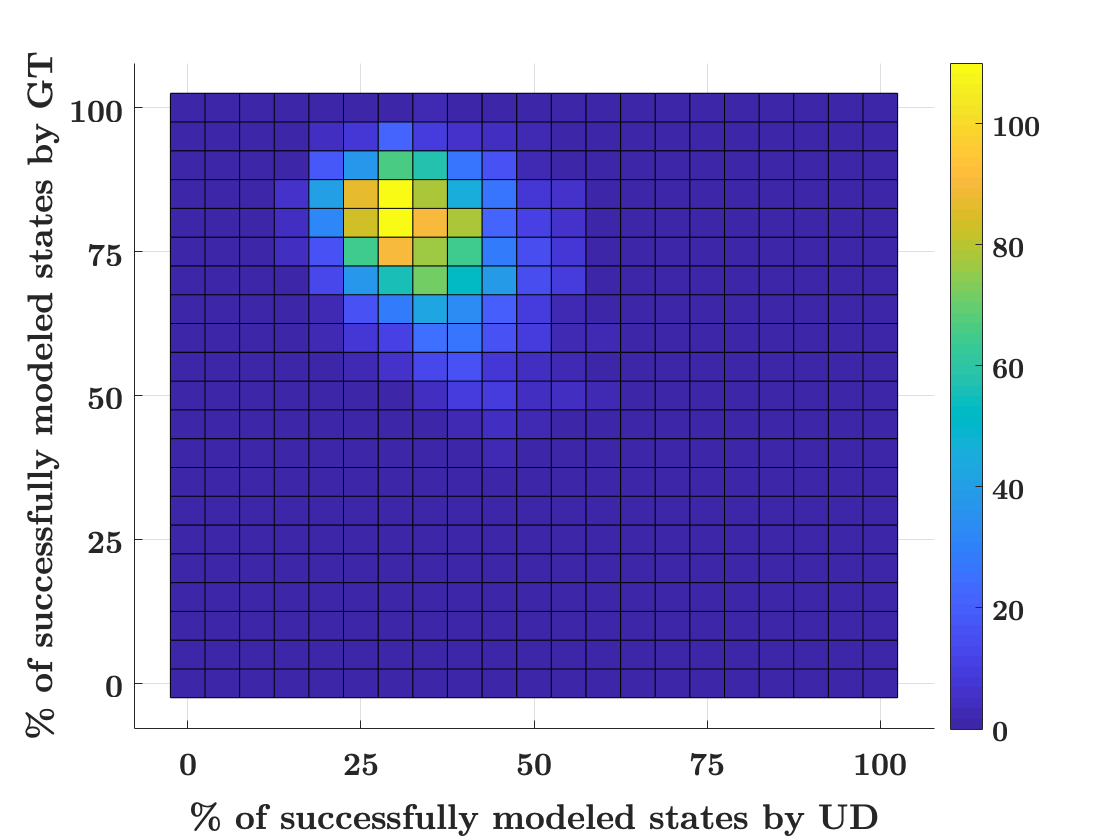}
		\caption{Color map showing the number of drivers whose x\% of the visited states are successfully modeled by the UD policy and y\% by the c-DQN-based GT policy. x and y percentages are given in the horizontal and vertical axes, respectively. }
		\label{fig:color3c}
     \end{subfigure}
        \caption{Comparison results for $n_{limit} = 3$ and $RL_{method}=c-DQN$ (US101)}
        \label{fig:res3c}
\end{figure*}

Performance of the GT policies, based on c-DQN, in terms of modeling drive behavior, and the difference between the GT and UD policy performances are given in Fig. \ref{fig:policy3c} and \ref{fig:dif3c}, respectively. The dramatic improvement in modeling percentages compared to the previous two cases, presented in subsections \textit{a)} and \textit{b)}, is a result of continuous, instead of discrete, observations. A color map similar to the one presented in Fig. \ref{fig:color3d}, but for the case where c-DQN is employed, instead of DQN, is also created and shown in Fig. \ref{fig:color3c}. Compared to Fig. \ref{fig:color3d}, the color cluster is further away from the x=y line which also emphasizes the dramatic improvement over the positive performance difference of GT policies over the UD policy. 

\medskip
\noindent \textit{d) $RL_{method}:$c-DQN, $n_{limit}=5$}
\smallskip

$aMAE = 0.62$ and $rMAE = 1.47$, for this model-data comparison.

\begin{figure*}
     \centering
     \begin{subfigure}[t]{0.49\textwidth}
        \centering
		\includegraphics[width=\textwidth]{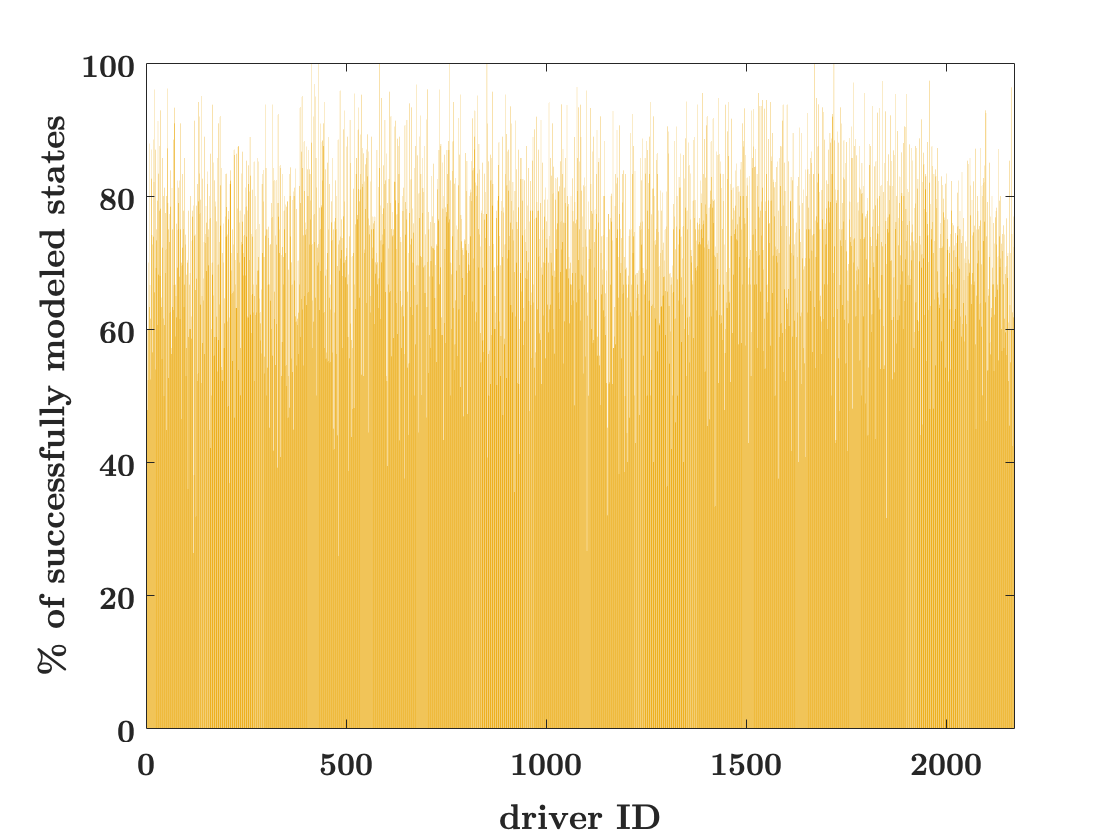}
		\caption{Percentages of successfully modeled states by the GT policies obtained through c-DQN, for each driver. Each vertical line belongs to an individual driver.}
		\label{fig:policy5c}
     \end{subfigure}
     \hfill
     \begin{subfigure}[t]{0.49\textwidth}
        \centering
		\includegraphics[width=\textwidth]{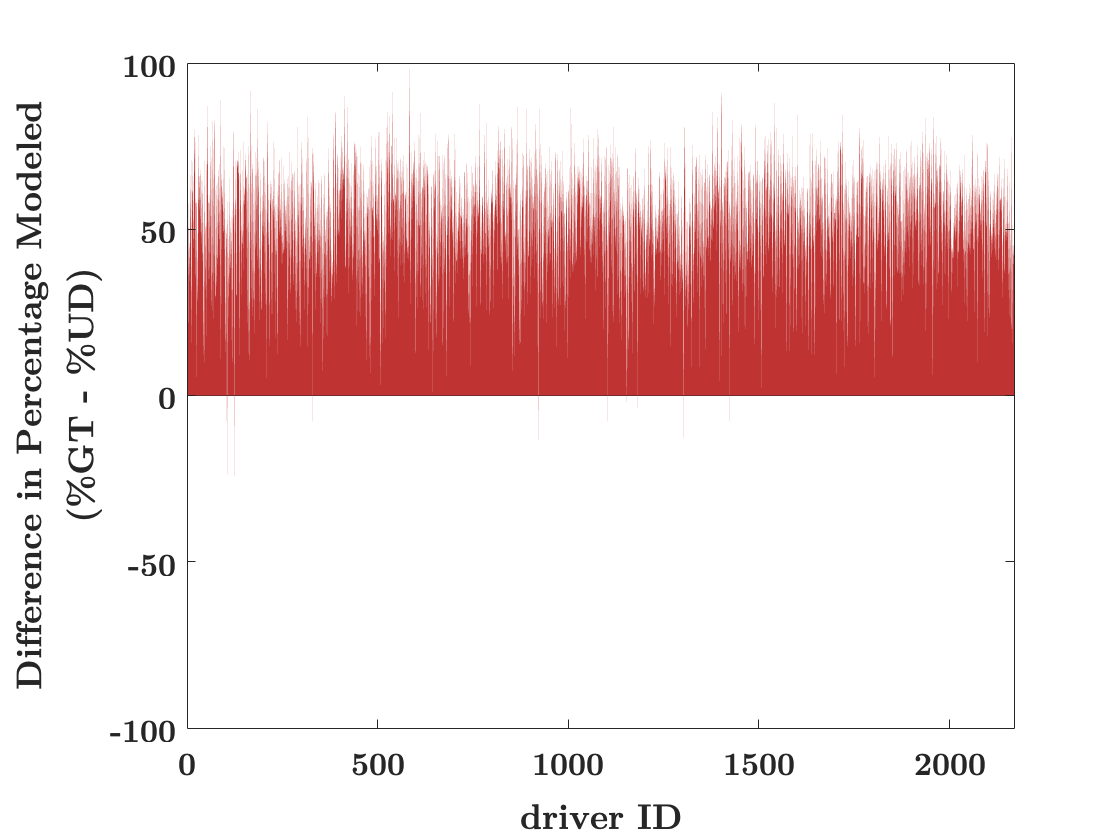}
		\caption{Differences in the percentages of the successfully modeled states of each driver, between the c-DQN-based GT policies and the UD policy.}
		\label{fig:dif5c}
     \end{subfigure}
     \hfill
     \begin{subfigure}[t]{0.49\textwidth}
        \centering
		\includegraphics[width=\textwidth]{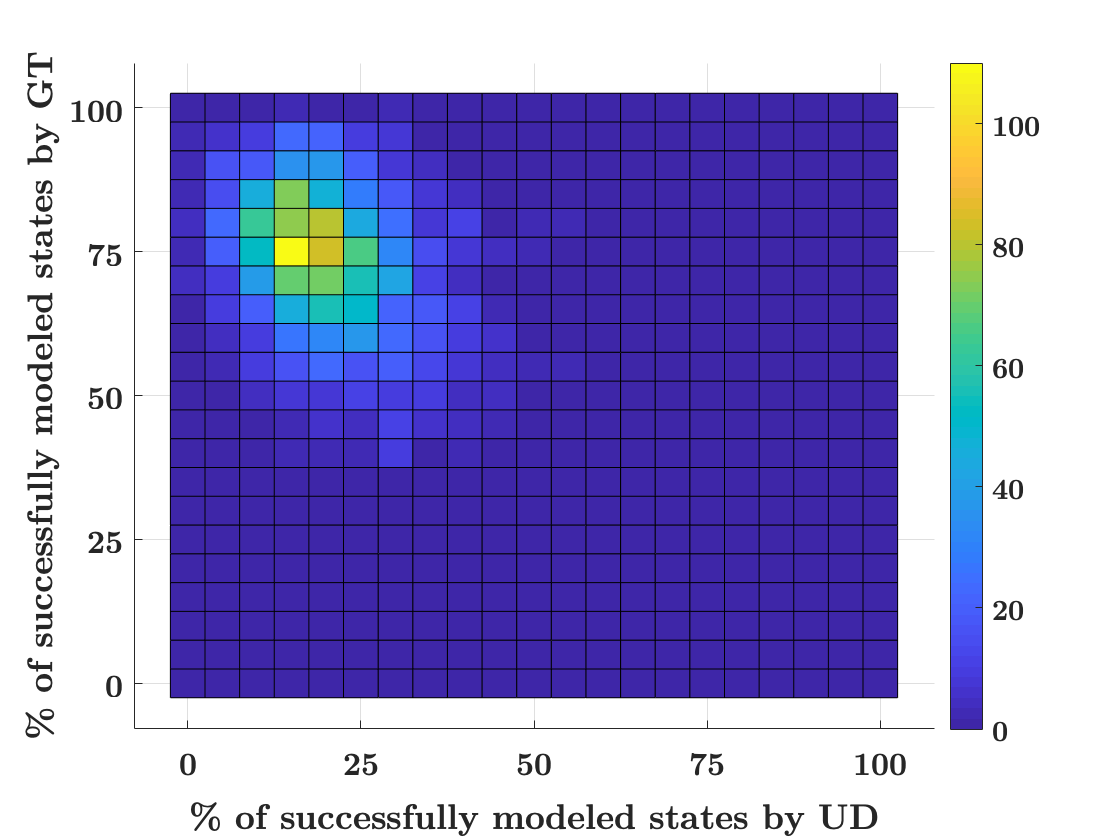}
		\caption{Color map showing the number of drivers whose x\% of the visited states are successfully modeled by the UD policy and y\% by the c-DQN-based GT policy. x and y percentages are given in the horizontal and vertical axes, respectively. }
		\label{fig:color5c}
     \end{subfigure}
        \caption{Comparison results for $n_{limit} = 5$ and $RL_{method}=c-DQN$ (US101)}
        \label{fig:res5c}
\end{figure*}

Fig. \ref{fig:policy5c} shows the percentages of successfully modeled driver behavior, for each of the 2168 drivers, by the c-DQN-based GT policies. Moreover, Fig. \ref{fig:dif5c} shows the difference between the successfully modeled visited state percentages of each driver by the GT policies and the UD policy. Fig. \ref{fig:color5c} presents a color map similar to the one given in Fig. \ref{fig:color3c}, but this time for the case of $n_{limit}=5$. Compared to the previous case, presented in subsection \textit{c)}, these figures, Figs. \ref{fig:policy5c}-c, show that the positive performance difference between the GT and UD policies improved. The main reason for this improvement is the increase $n_{limit}$ value, which corresponds to an increased K-S test power. 

\subsubsection{Model Validation with I-80 Data}
In addition to US101, highway I-80 data \cite{I80}, is also used to test the validity of the proposed GT policies. For this test, I-80 data collected between 5.00-5.15 PM is used, which contains 1835 drivers. 

\medskip
\noindent \textit{a)$RL_{method}:$DQN, $n_{limit}=3$}
\smallskip

For this model-data comparison, $aMAE = 0.27$ and $rMAE = 1.62$.

\begin{figure*}
     \centering
     \begin{subfigure}[t]{0.49\textwidth}
        \centering
		\includegraphics[width=\textwidth]{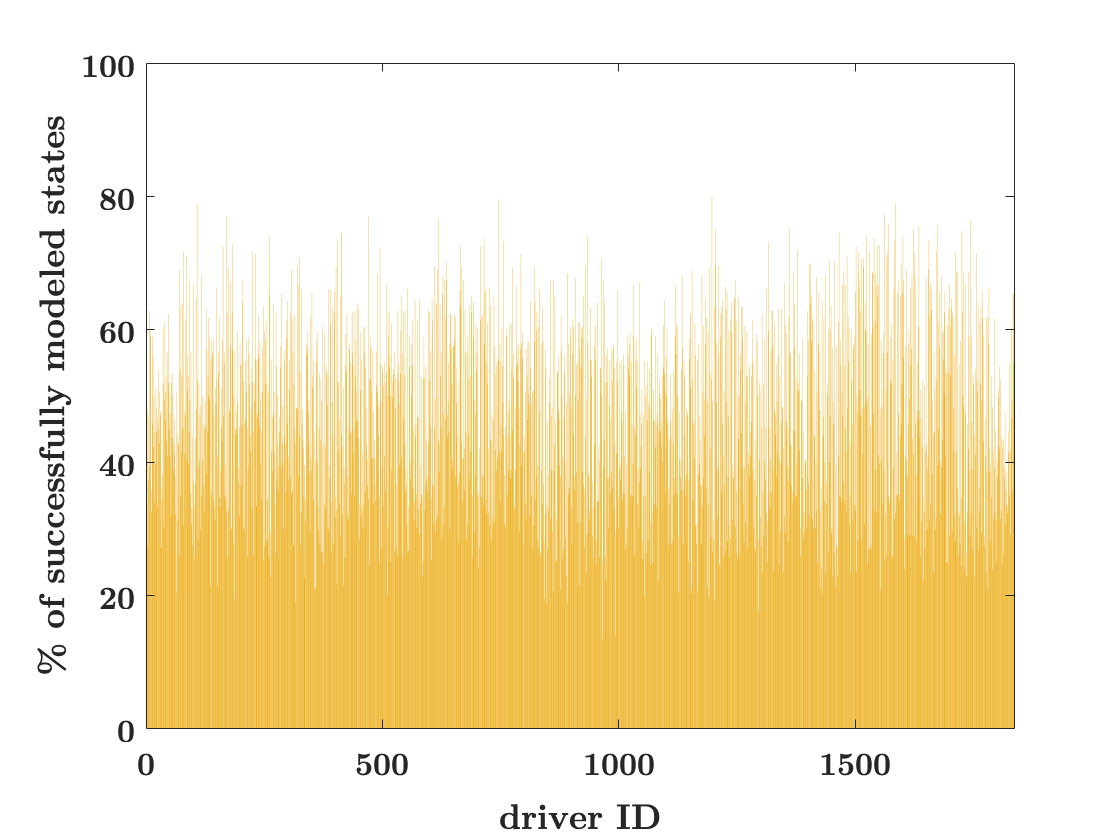}
		\caption{Percentages of successfully modeled states by the GT policies obtained through DQN, for each driver. Each vertical line belongs to an individual driver.}
		\label{fig:policy3di}
     \end{subfigure}
     \hfill
     \begin{subfigure}[t]{0.49\textwidth}
        \centering
		\includegraphics[width=\textwidth]{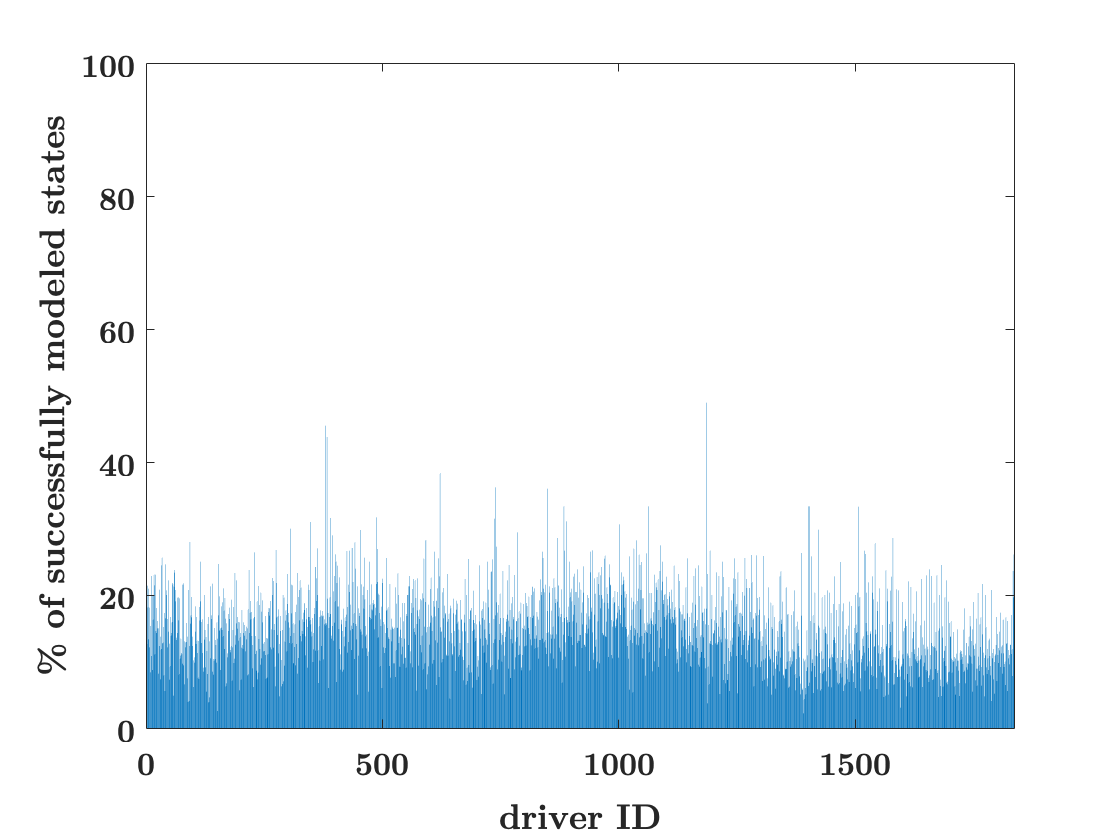}
		\caption{Percentages of successfully modeled states by the UD policy, for each driver. Each vertical line belongs to an individual driver..}
		\label{fig:dumb3di}
     \end{subfigure}
     \hfill
     \begin{subfigure}[t]{0.49\textwidth}
        \centering
		\includegraphics[width=\textwidth]{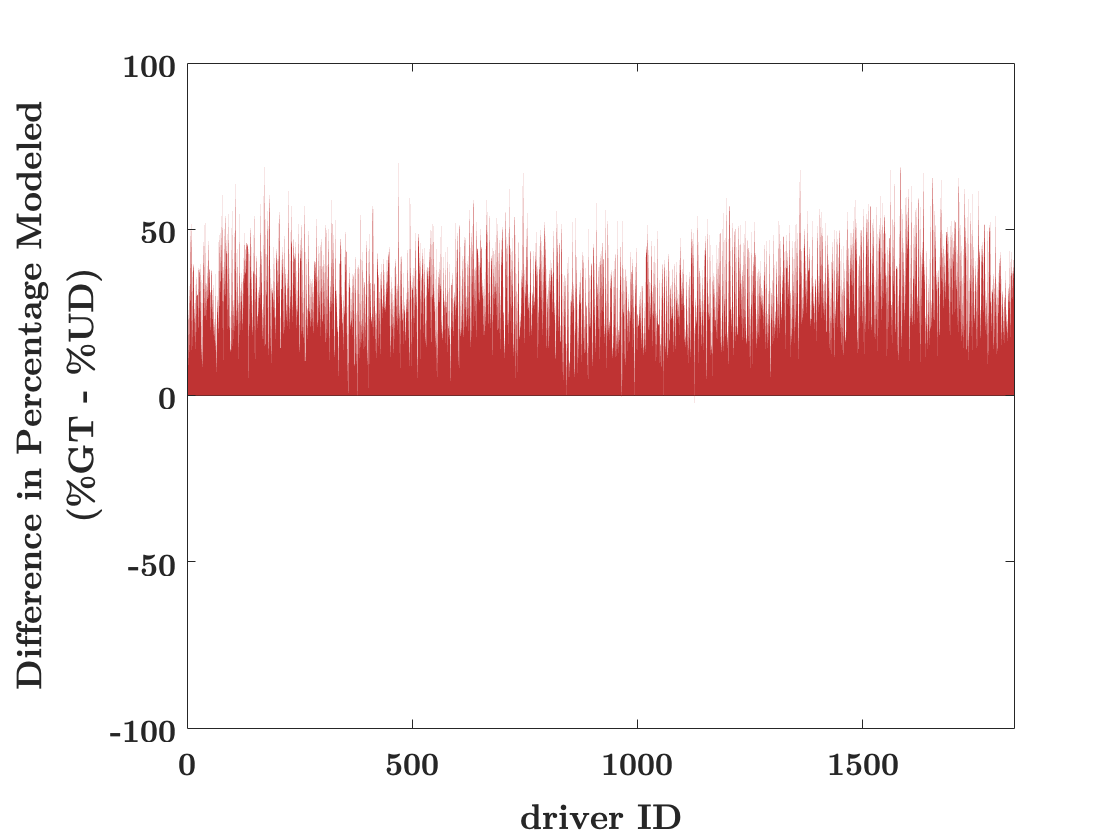}
		\caption{Differences in the percentages of the successfully modeled states of each driver, between the DQN-based GT policies and the UD policy.}
		\label{fig:dif3di}
     \end{subfigure}
     \hfill
     \begin{subfigure}[t]{0.49\textwidth}
        \centering
		\includegraphics[width=\textwidth]{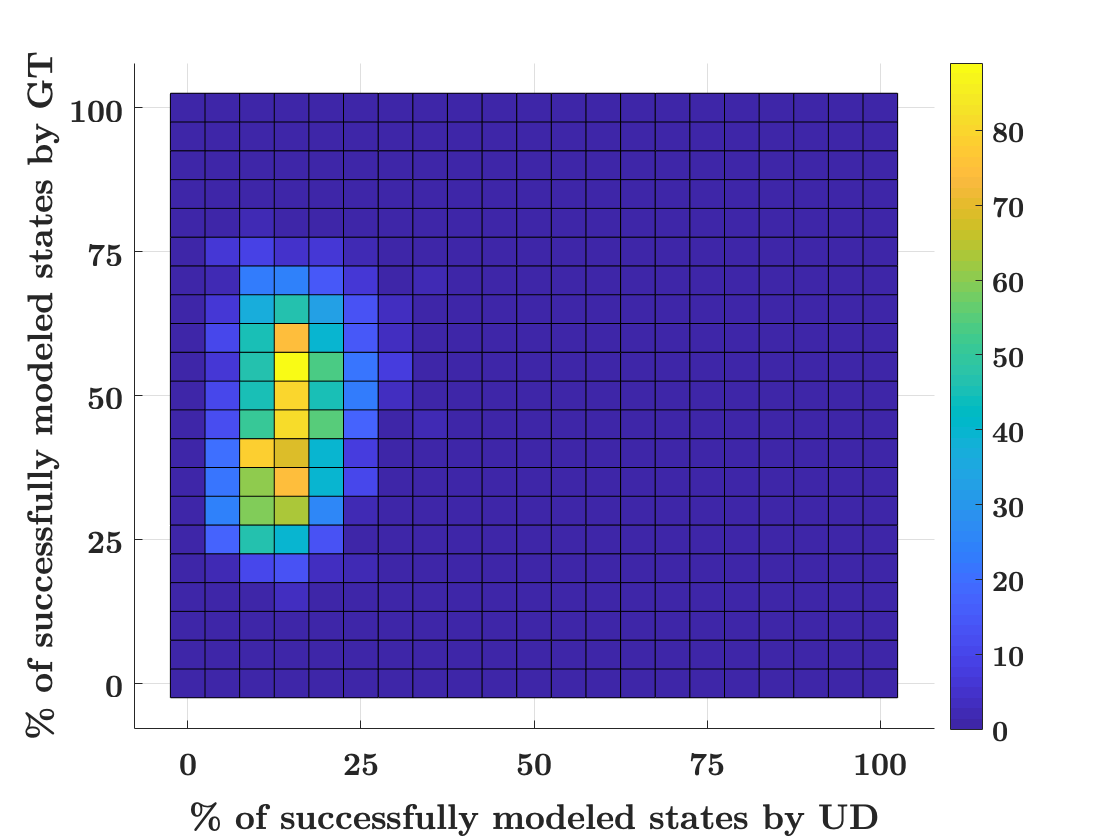}
		\caption{Color map showing the number of drivers whose x\% of the visited states are successfully modeled by the UD policy and y\% by the DQN-based GT policy. x and y percentages are given in the horizontal and vertical axes, respectively. }
		\label{fig:color3di}
     \end{subfigure}
        \caption{Comparison results for $n_{limit} = 3$ and $RL_{method}=DQN$ (I80)}
        \label{fig:res3di}
\end{figure*}

For each 1835 human drivers in the dataset, percentages of visited states whose policies are succesfully modeled by the GT policies and the UD policy are presented in Figs. \ref{fig:policy3di} and \ref{fig:dumb3di}, respectively. These figures show that the GT policies model human drivers considerably better than the UD policies. The difference between the percentages of of the successfully modeled policies by the GT policies and the UD policy is given in Fig. \ref{fig:dif3di}. When compared with Fig. \ref{fig:dif3d}, Fig. \ref{fig:dif3di} shows that the proposed GT policies have a higher percentage of success rate in modeling drivers in I80 compared to the ones in US101. In Fig. \ref{fig:color3di} x and y axes (horizontal and vertical) show the percentages of the successfully modeled states by the UD and GT policies, respectively. The color cluster in Fig. \ref{fig:color3di} being above the x=y line shows that the GT policies perform better than the UD policy. Compared to Fig. \ref{fig:color3d}, Fig. \ref{fig:color3di} shows that the positive performance difference between the GT and UD policies is larger for I80 data compared to that of US101. 

\medskip
\noindent \textit{b)$RL_{method}:$DQN, $n_{limit}=5$}
\smallskip

$aMAE = 0.22$ and $rMAE = 1.61$, for this model-data comparison.

\begin{figure*}
     \centering
     \begin{subfigure}[t]{0.49\textwidth}
        \centering
		\includegraphics[width=\textwidth]{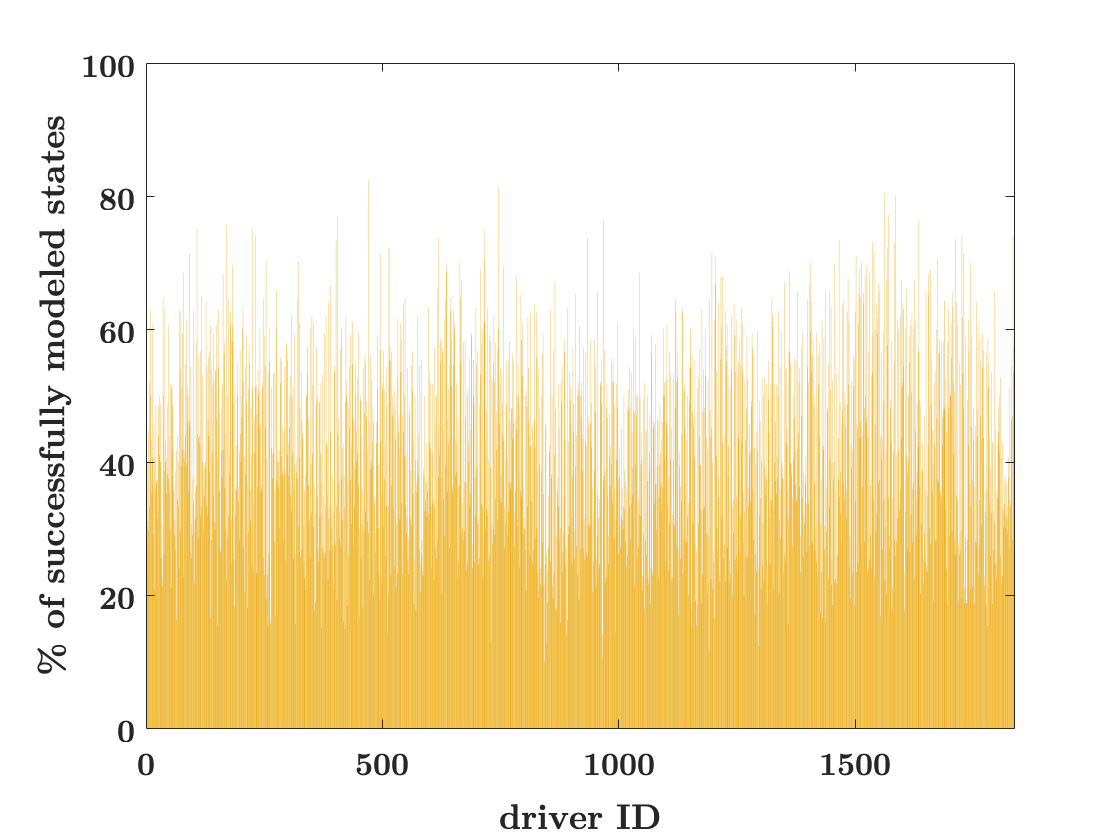}
		\caption{Percentages of successfully modeled states by the GT policies obtained through DQN, for each driver. Each vertical line belongs to an individual driver.}
		\label{fig:policy5di}
     \end{subfigure}
     \hfill
     \begin{subfigure}[t]{0.49\textwidth}
        \centering
		\includegraphics[width=\textwidth]{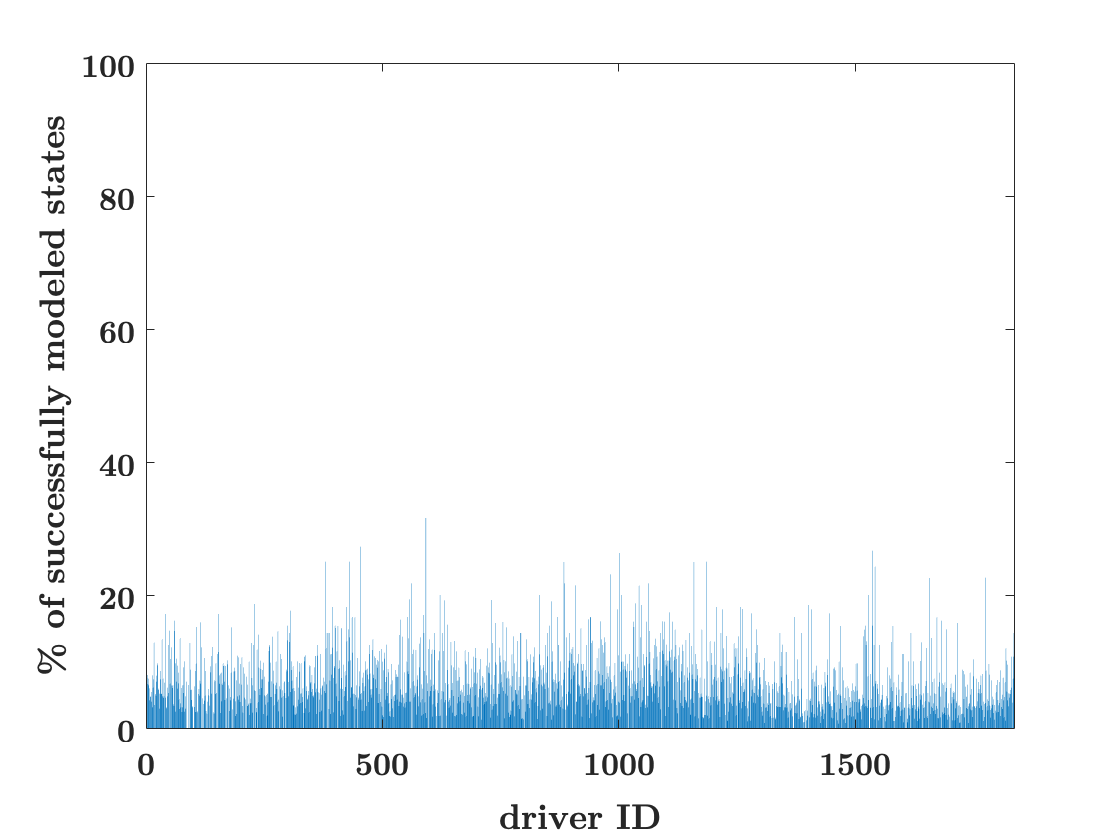}
		\caption{Percentages of successfully modeled states by the UD policy, for each driver. Each vertical line belongs to an individual driver..}
		\label{fig:dumb5di}
     \end{subfigure}
     \hfill
     \begin{subfigure}[t]{0.49\textwidth}
        \centering
		\includegraphics[width=\textwidth]{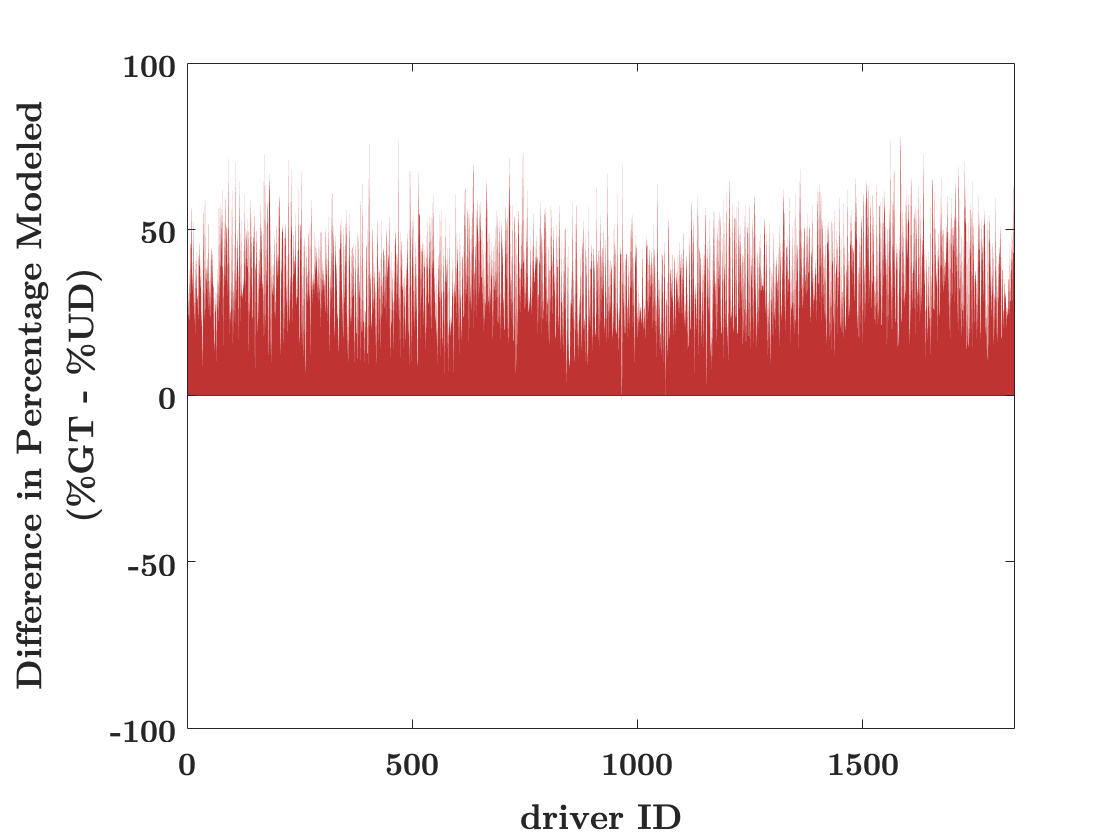}
		\caption{Differences in the percentages of the successfully modeled states of each driver, between the DQN-based GT policies and the UD policy.}
		\label{fig:dif5di}
     \end{subfigure}
     \hfill
     \begin{subfigure}[t]{0.49\textwidth}
        \centering
		\includegraphics[width=\textwidth]{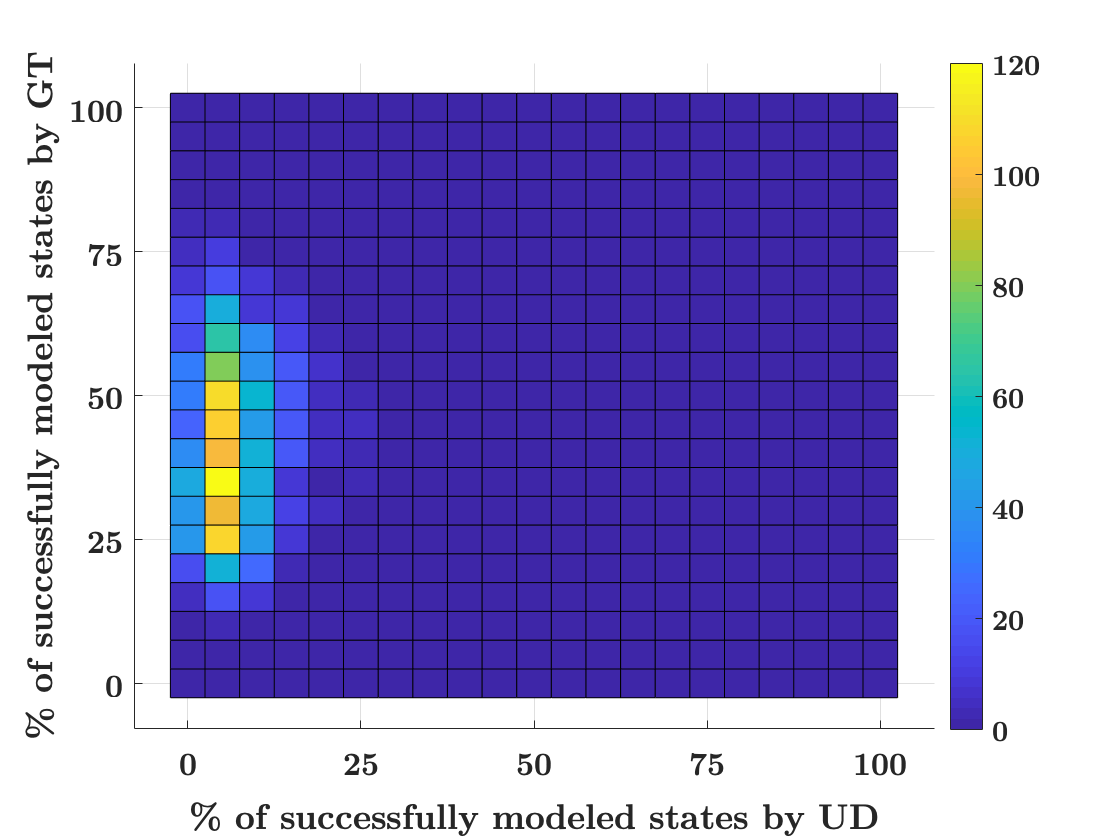}
		\caption{Color map showing the number of drivers whose x\% of the visited states are successfully modeled by the UD policy and y\% by the DQN-based GT policy. x and y percentages are given in the horizontal and vertical axes, respectively. }
		\label{fig:color5di}
     \end{subfigure}
        \caption{Comparison results for $n_{limit} = 5$ and $RL_{method}=DQN$ (I80)}
        \label{fig:res5di}
\end{figure*}

Driver behavior modeling performance of the GT and UD policies are given in Fig. \ref{fig:policy5di} and \ref{fig:dumb5di}, respectively. The difference between these policies, in terms of successfully modeled state percentages, are given in Fig \ref{fig:dif5di}. A color map, similar to Fig. \ref{fig:color5d}, is also shown in Fig. \ref{fig:color5di}. Compared to Figs. \ref{fig:policy3di}-d, Figs. \ref{fig:policy5di}-d show that with the increase in the $n_{limit}$ value, i.e. with the increase in the test power, the difference between the GT and the UD policies becomes more clear in terms of human driver behavior modeling performance.

\medskip
\noindent \textit{c) $RL_{method}:$c-DQN, $n_{limit}=3$}
\smallskip

$aMAE = 0.64$ and $rMAE = 1.60$, for this model-data comparison.

\begin{figure*}
     \centering
     \begin{subfigure}[t]{0.49\textwidth}
        \centering
		\includegraphics[width=\textwidth]{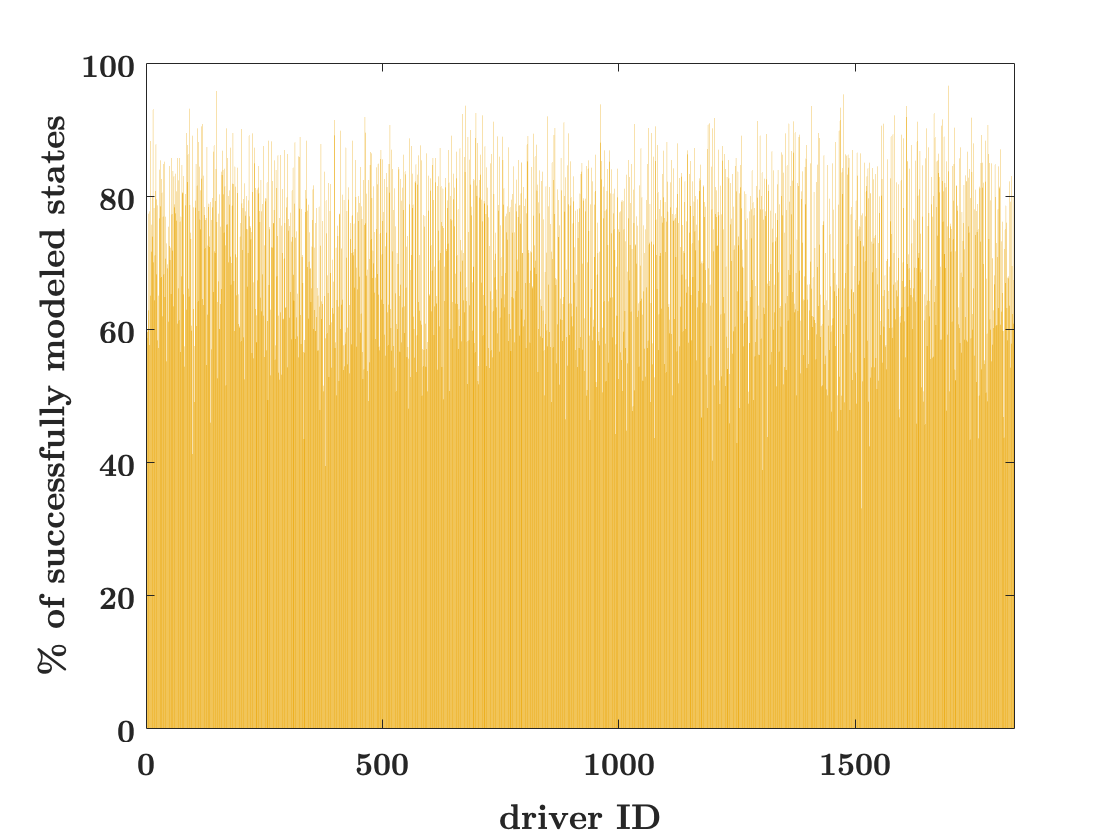}
		\caption{Percentages of successfully modeled states by the GT policies obtained through c-DQN, for each driver. Each vertical line belongs to an individual driver.}
		\label{fig:policy3ci}
     \end{subfigure}
     \hfill
     \begin{subfigure}[t]{0.49\textwidth}
        \centering
		\includegraphics[width=\textwidth]{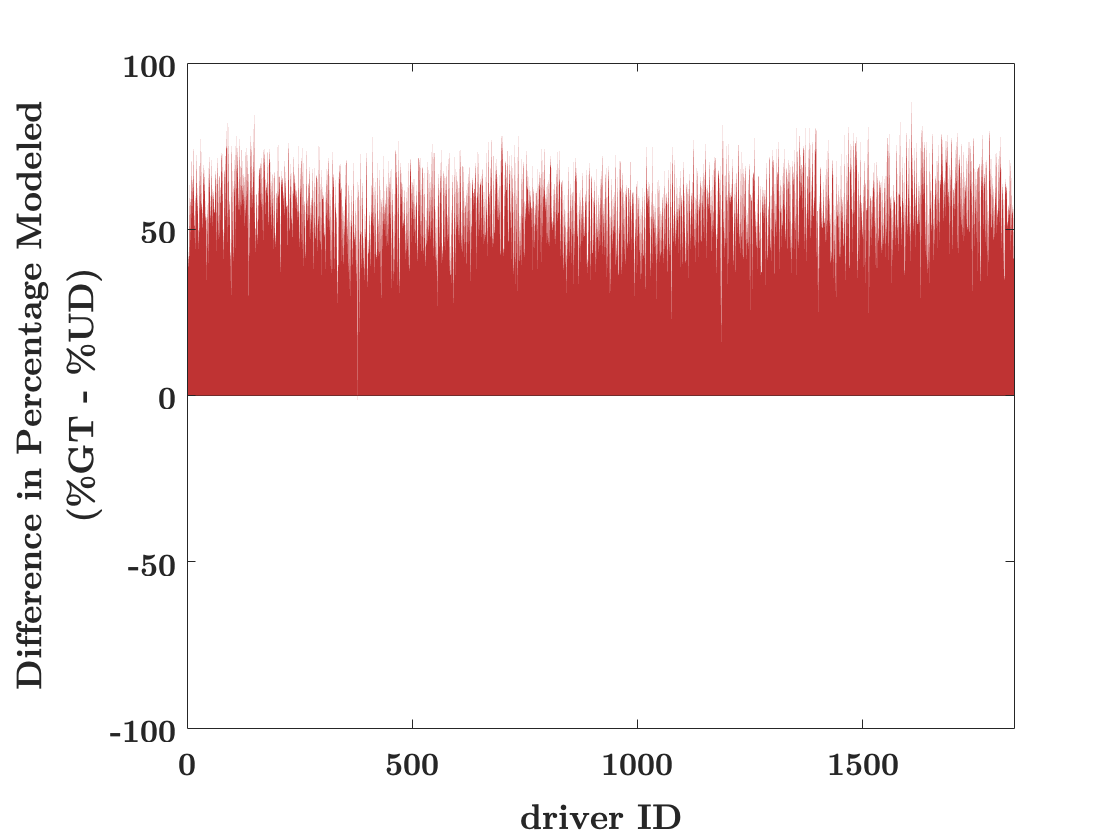}
		\caption{Differences in the percentages of the successfully modeled states of each driver, between the c-DQN-based GT policies and the UD policy.}
		\label{fig:dif3ci}
     \end{subfigure}
     \hfill
     \begin{subfigure}[t]{0.49\textwidth}
        \centering
		\includegraphics[width=\textwidth]{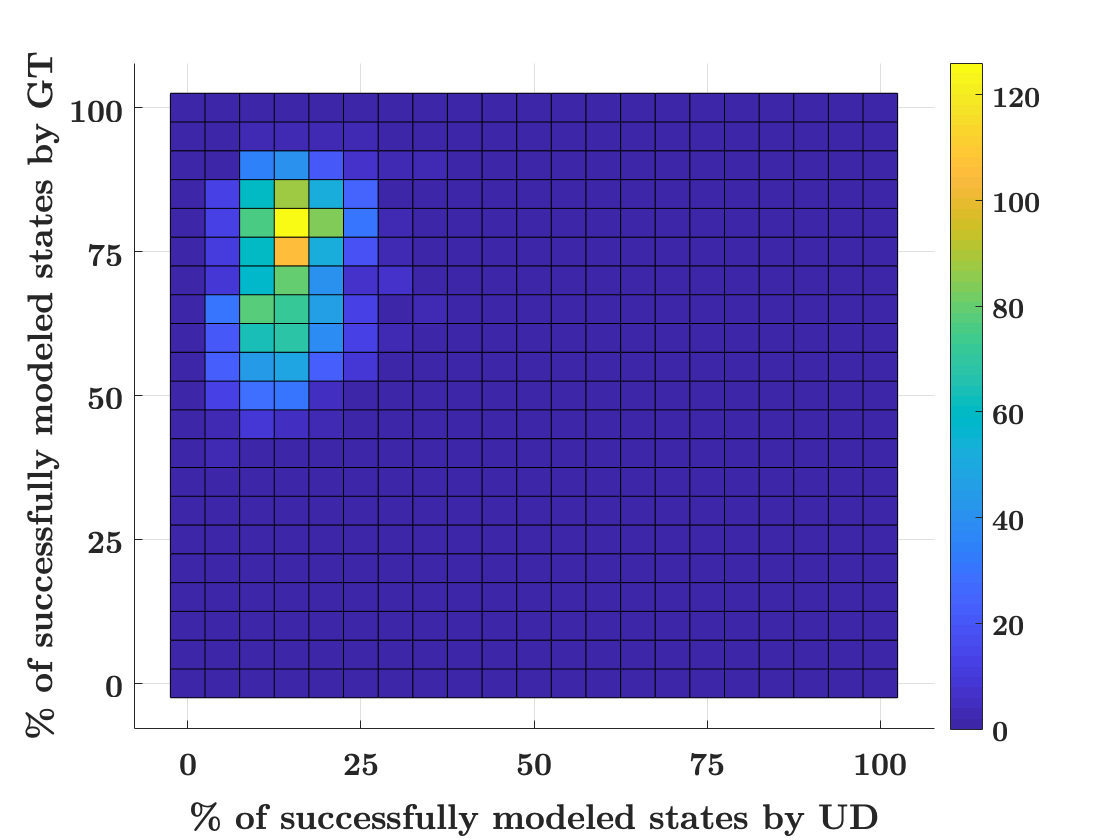}
		\caption{Color map showing the number of drivers whose x\% of the visited states are successfully modeled by the UD policy and y\% by the c-DQN-based GT policy. x and y percentages are given in the horizontal and vertical axes, respectively. }
		\label{fig:color3ci}
     \end{subfigure}
        \caption{Comparison results for $n_{limit} = 3$ and $RL_{method}=c-DQN$ (I80)}
        \label{fig:res3ci}
\end{figure*}

In Fig. \ref{fig:policy3ci}, the performance of the GT policies in terms of modeling driver performance is shown and in Fig. \ref{fig:dif3ci}, the difference between percentage of states whose policies are successfully modeled by the GT policies and by the UD policy is given. Fig. \ref{fig:color3ci} shows a color map similar to the one presented in Fig. \ref{fig:color3di}, but for the case where c-DQN, instead of DQN, is employed. Compared to the previous two cases, the results of which are provided in Figs. \ref{fig:policy3di}-d and \ref{fig:policy5di}-d, Figs. \ref{fig:policy3ci}-c show a dramatic improvement on the modeling capability of the GT policies, thanks to the continuous observation space. Furthermore, a comparison between Fig. \ref{fig:policy3c}-c and Fig. \ref{fig:policy3ci}-c shows that proposed GT policies' performance advantage over UD is more pronounced for I80 compared to US101. 

\medskip
\noindent \textit{d) $RL_{method}:$c-DQN, $n_{limit}=5$}
\smallskip

For this model-data comparison, $aMAE = 0.52$ and $rMAE = 1.53$.

\begin{figure*}
     \centering
     \begin{subfigure}[t]{0.49\textwidth}
        \centering
		\includegraphics[width=\textwidth]{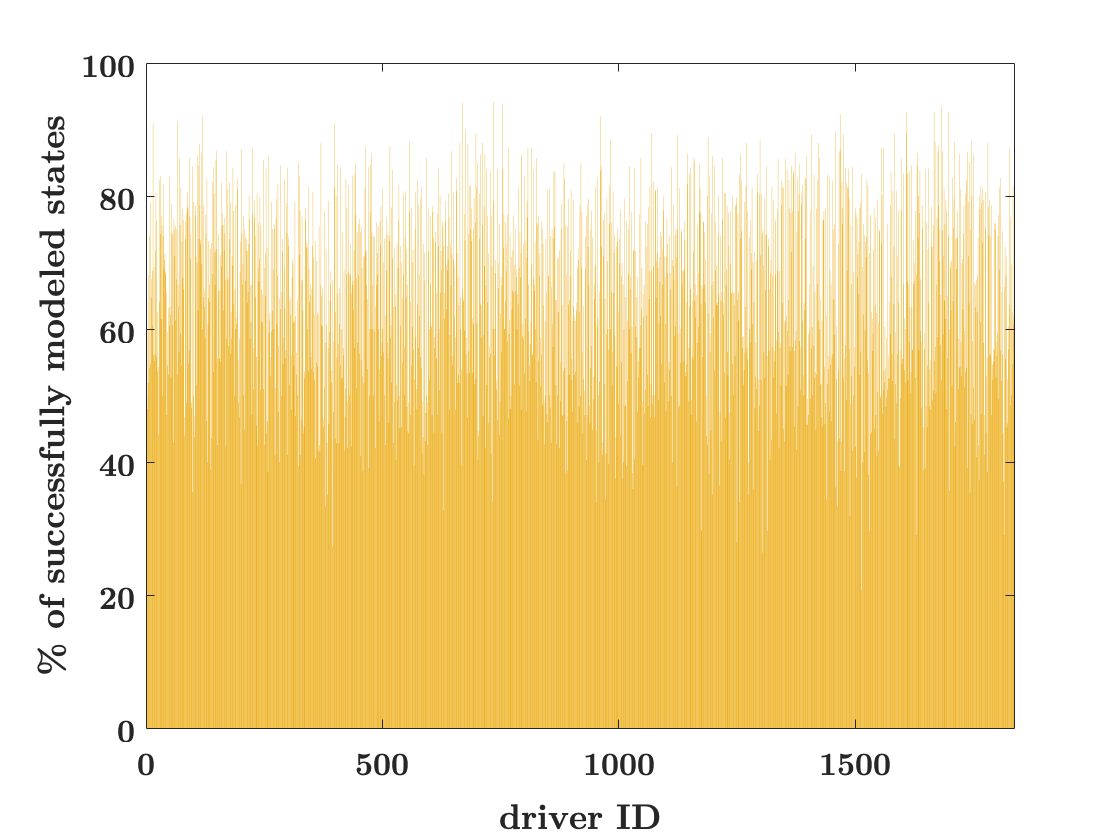}
		\caption{Percentages of successfully modeled states by the GT policies obtained through c-DQN, for each driver. Each vertical line belongs to an individual driver.}
		\label{fig:policy5ci}
     \end{subfigure}
     \hfill
     \begin{subfigure}[t]{0.49\textwidth}
        \centering
		\includegraphics[width=\textwidth]{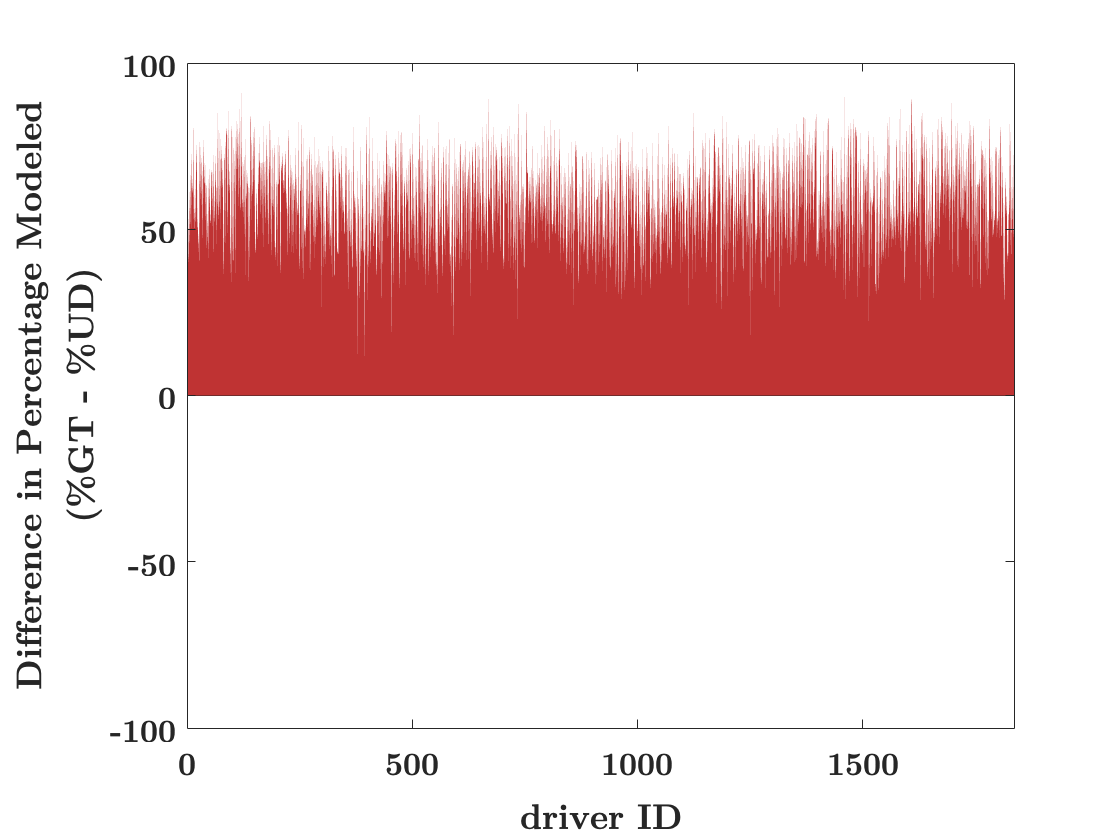}
		\caption{Differences in the percentages of the successfully modeled states of each driver, between the c-DQN-based GT policies and the UD policy.}
		\label{fig:dif5ci}
     \end{subfigure}
     \hfill
     \begin{subfigure}[t]{0.49\textwidth}
        \centering
		\includegraphics[width=\textwidth]{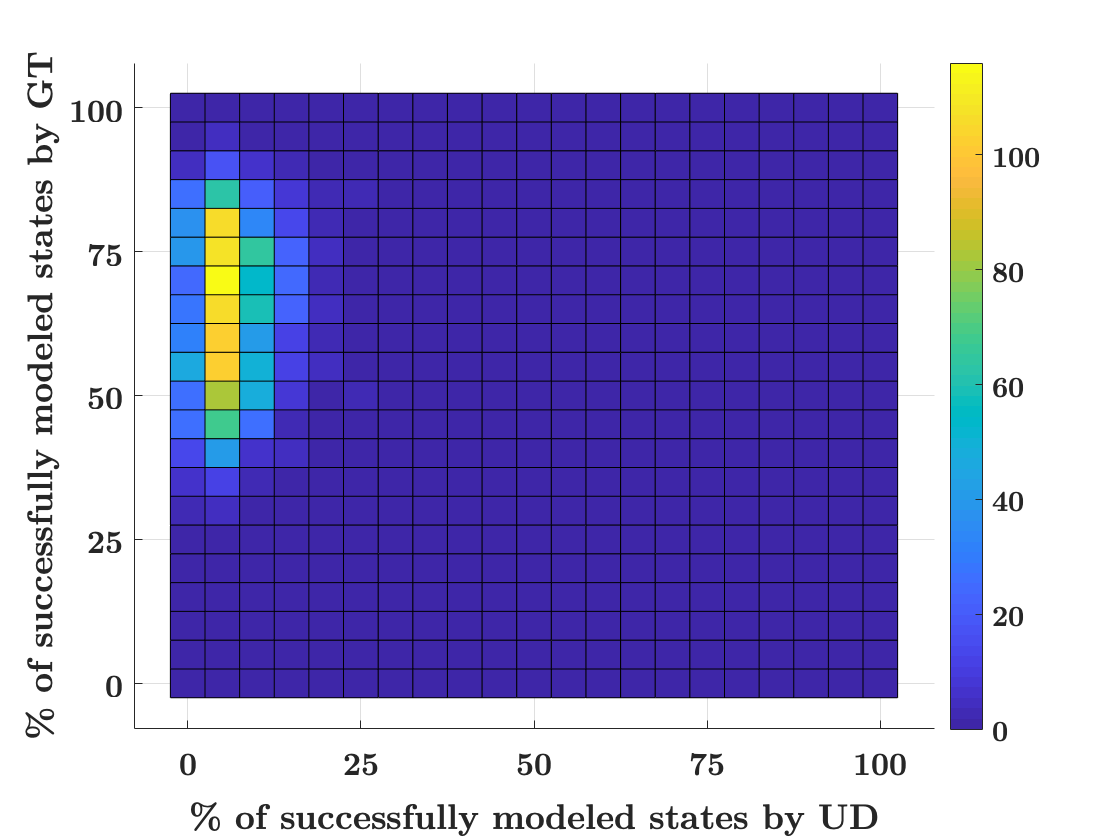}
		\caption{Color map showing the number of drivers whose x\% of the visited states are successfully modeled by the UD policy and y\% by the c-DQN-based GT policy. x and y percentages are given in the horizontal and vertical axes, respectively. }
		\label{fig:color5ci}
     \end{subfigure}
        \caption{Comparison results for $n_{limit} = 5$ and $RL_{method}=c-DQN$ (I80)}
        \label{fig:res5ci}
\end{figure*}

Performance of the GT policies in terms of modeling human driver behavior is presented in Fig. \ref{fig:policy5ci}. Furthermore, for each of the 1835 drivers, the difference between percentages of successfully modeled states by the GT policies and the UD policy is shown in Fig. \ref{fig:dif5ci}. A color map, similar to the one presented in Fig. \ref{fig:color3ci} is given in Fig. \ref{fig:color5ci}. Compared to Figs. \ref{fig:policy3ci}-c, Figs. \ref{fig:policy5ci}-c show that the increase in $n_{limit}$, which increases the K-S test power, demonstrates the positive difference between the modeling capability of the GT and UD policies more clearly. 

\begin{remark}
US101 data is used only to determine the observation and action set boundaries. It is not used to \textit{train} the GT driver models. Therefore, the GT policies are not obtained by fitting the model parameters to the data. However, since this data is used to set the observation-action space boundaries, it still affected, albeit indirectly, the obtained models. To test the resulting GT policies with data that is not used in any way to obtain these policies, additional model-validation tests are conducted with the I80 data. To summarize, although the US101 data is not used to train the models, and therefore \textit{overfitting} is not a concern, additional validation tests are conducted with the I80 data for further assurance of the validity of the GT models. 
\end{remark}

\subsubsection{Summary of Results}
For a more clear presentation of the statistical analysis results, Table I and Table II are given, which provide a summary of each K-S test conducted with different parameters.

\begin{table*}
\centering
\caption{Average performance of both the GT policies and the UD policy in terms of modeling human driver behaviors for US101 data.}
\begin{tabular}{|c||c|c|c|c|c|} 
\hline
\multicolumn{1}{|c}{}                    &                                                                                       & \begin{tabular}[c]{@{}c@{}}\textbf{Mean \% of states whose policies are successfully}\\ \textbf{modeled}\textbf{by GT policies} \end{tabular} & \begin{tabular}[c]{@{}c@{}}\textbf{Mean \% Difference in~modeled percentages }\\\textbf{(GT\% - UD\%)}\end{tabular} & \multicolumn{1}{l|}{\textbf{aMAE} } & \multicolumn{1}{l|}{\textbf{rMAE} }  \\ 
\hline
\multirow{2}{*}{\rotatebox[origin=c]{90}{$n_{limit}=3$}} & \begin{tabular}[c]{@{}c@{}}\textbf{}~\\\textbf{c-DQN Policies}\\\textbf{~}\end{tabular}         & 76.73\%                                                                                    & 42.71\%                                                                                                             & 0.71                                & 1.51                                 \\ 
\cline{2-6}
                                         & \begin{tabular}[c]{@{}c@{}}\textbf{}~\\\textbf{DQN Policies}\\\textbf{~}\end{tabular} & 40.24\%                                                                                    & 6.24\%                                                                                                              & 0.50                                & 1.56                                 \\ 
\hline
\multirow{2}{*}{\rotatebox[origin=c]{90}{$n_{limit}=5$}} & \begin{tabular}[c]{@{}c@{}}\textbf{}~\\\textbf{c-DQN Policies}\\\textbf{~}\end{tabular}          & 72.72\%                                                                                    & 52.21\%                                                                                                             & 0.62                                & 1.42                                 \\ 
\cline{2-6}
                                         & \begin{tabular}[c]{@{}c@{}}\textbf{}~\\\textbf{DQN Policies}\\\textbf{~}\end{tabular} & 34.75\%                                                                                    & 14.26\%                                                                                                             & 0.42                                & 1.54                                 \\
\hline
\end{tabular}
\end{table*}

\begin{table*}
\centering
\caption{Average performance of both the GT policies and the UD policy in terms of modeling human driver behaviors for I-80 data.}
\begin{tabular}{|c||c|c|c|c|c|} 
\hline
\multicolumn{1}{|c}{}                    &                                                                                       & \begin{tabular}[c]{@{}c@{}}\textbf{Mean \% of states whose policies are successfully}\\ \textbf{modeled}\textbf{by GT policies} \end{tabular} & \begin{tabular}[c]{@{}c@{}}\textbf{Mean \% Difference in~modeled percentages }\\\textbf{(GT\% - UD\%)}\end{tabular} & \multicolumn{1}{l|}{\textbf{aMAE} } & \multicolumn{1}{l|}{\textbf{rMAE} }  \\ 
\hline
\multirow{2}{*}{\rotatebox[origin=c]{90}{$n_{limit}=3$}} & \begin{tabular}[c]{@{}c@{}}\textbf{}~\\\textbf{c-DQN Policies}\\\textbf{~}\end{tabular}         & 71.52\%                                                                                    & 56.65\%                                                                                                             & 0.62                                & 1.54                                 \\ 
\cline{2-6}
                                         & \begin{tabular}[c]{@{}c@{}}\textbf{}~\\\textbf{DQN Policies}\\\textbf{~}\end{tabular} & 46.41\%                                                                                    & 31.55\%                                                                                                              & 0.27                              & 1.63                                 \\ 
\hline
\multirow{2}{*}{\rotatebox[origin=c]{90}{$n_{limit}=5$}} & \begin{tabular}[c]{@{}c@{}}\textbf{}~\\\textbf{c-DQN Policies}\\\textbf{~}\end{tabular}          & 64.46\%                                                                                    & 58.12\%                                                                                                             & 0.51                               & 1.48                                 \\ 
\cline{2-6}
                                         & \begin{tabular}[c]{@{}c@{}}\textbf{}~\\\textbf{DQN Policies}\\\textbf{~}\end{tabular} & 41.85\%                                                                                    & 35.52\%                                                                                                             & 0.22                                & 1.62                                 \\
\hline
\end{tabular}
\end{table*}

Table I presents the results obtained using US101 data. Percentage of visited states whose policies are successfully modeled by the GT policies, averaged over all drivers, is given in the first column. The difference between GT and UD policy performances are given in the second column. Average mean errors for the policies that passed the K-S test, aMAE, and for the ones that failed to pass the test, rMAE, are provided in the third and fourth columns, respectively. The results show that the GT policies developed using DQN model driver behavior better than the UD policy. The advantage of GT becomes dramatically more pronounced when c-DQN is employed, reaching up to 52\% lead over UD. Moreover, it is seen that as the power of the test increases with the increase in the value of $n_{limit}$, the difference between GT policies and the UD policy becomes more clear. The effect of the increase in $n_{limit}$ also shows itself in aMAE and rMAE values: As the test power increases, these values decrease, meaning that the test's ability to distinguish the model policies, either GT or UD, from the data-based policies increases. This increases the confidence in test results. 

Statistical analysis summary for the I80 data is provided similarly in Table 2. Similar conclusions as stated for Table 1 can be drawn for the results shown in Table 2. The main difference is that the power of the GT policies are more pronounced here compared to the US101 data. The difference does not stem from a dramatic improvement of the success of the GT policies but a large drop in the predictive power of the UD policies for this dataset. One conclusion can be that the driver reactions given in I80 dataset are much harder to model using a UD model compared to US101. 

Finally, the driver policies being modeled with up to 77\% success rate for US101, and 72\% for I80 show that the proposed GT models are strong candidates for a high-fidely representation of real human driver reactions in highways. 

\section{Summary}
In this paper, a modeling framework combining a game theory (GT) concept named level-k reasoning and a deep reinforcement learning method called deep Q-learning, is proposed. Both the discrete (DQN) and continuous (c-DQN) versions of the reinforcement learning method are tested in the framework, for modeling human driver behaviors in highway traffic. It is observed that, compared to earlier similar studies, the crash rates of the proposed driver models are significantly smaller and thus more realistic. Reduced crash rates are attributed to the significantly enlarged observation space and therefore reducing the number of blind spots for the driver. 

For evaluating the predictive power of the GT models, two independent traffic data sets, obtained from highways US101 and I80, are used. GT models and the driver policies derived from processing these two raw traffic data sets are compared using Kolmogorov-Smirnov (K-S) goodness of fit test. The results of the statistical analysis conducted using the K-S test shows that on average, proposed GT policies can  model up to 77\% the driver policies obtained from the US101 dataset and 72\% of the policies from the I-80 dataset. Furthermore, GT policies performs 52\% better than a bencmark policy formed using a uniform distribution over the action space, for the US101 dataset, and 57\% better for the I-80 dataset.

\clearpage

\clearpage
\bibliographystyle{IEEEtran}\bibliography{biblo}

\end{document}